\documentclass[letter,useAMS,usenatbib]{mn2e}

\usepackage{color}
\usepackage{epsfig}
\usepackage{rotating}
\usepackage{multirow}
\usepackage{amsmath}
\usepackage{amssymb}

\newcommand*{\figref}[1]{Fig.~\ref{#1}}
\newcommand*{\tabref}[1]{Table~\ref{#1}}

\def\herschel{{\em Herschel}}
\def\spire{\herschel-SPIRE}
\def\pacs{\herschel-PACS}
\def\spitzer{{\em Spitzer}}
\def\planck{{\em Planck}}
\def\hfi{\planck-HFI}
\def\dd{\,\mathrm{d}}
\def\ld{\lambda/D}
\def\lod{\lambda_0/D}
\def\mic{$\umu$m}

\def\intnu{\int\limits_\nu}
\def\intbeam{\iint\limits_\mathrm{Beam}}

\def\srf{F(\nu)}
\def\smeas{\overline{S}_\mathrm{Meas}}
\def\spip{S_\mathrm{Pip}}
\def\ipip{I_\mathrm{PipE}}
\def\kmonp{K_\mathrm{MonP}}
\def\kcolp{K_\mathrm{ColP}}
\def\kmone{K_\mathrm{MonE}}
\def\kcole{K_\mathrm{ColE}}

\def\kuni{K_\mathrm{Uniform}}
\def\pmeas{P_\mathrm{Meas}}
\def\omeas{\Omega_\mathrm{Meas}}
\def\pmod{P_\mathrm{mod}}
\def\ppred{P_\mathrm{Pred}}
\def\opred{\Omega_\mathrm{Pred}}
\def\onorm{\Omega_\mathrm{norm}}
\def\oeff{\Omega_\mathrm{eff}}
\def\anep{\alpha_\mathrm{Nep}}
\def\nueff{\nu_\mathrm{eff}}

\title[Flux Calibration of far-IR \& sub-mm photometers]{Flux Calibration of Broadband Far Infrared and Submillimetre Photometric Instruments: Theory and Application to \herschel-SPIRE}

\author[M.~J.~Griffin et al.]{
M.~J.~Griffin,$^1$ 
C.~E.~North,$^1$  
B.~Schulz,$^{2,3}$ 
A.~Amaral-Rogers,$^1$ 
G.~Bendo,$^4$ 
J.~Bock,$^{2,5}$ 
\newauthor
L.~Conversi,$^6$ 
A.~Conley,$^7$ 
C.~D.~Dowell,$^5$ 
M.~Ferlet,$^8$ 
J.~Glenn,$^7$ 
T.~Lim,$^8$ 
C.~Pearson,$^{8,9}$ 
\newauthor
M.~Pohlen,$^{10}$ 
B.~Sibthorpe,$^{11}$ 
L.~Spencer,$^1$ 
B.~Swinyard,$^{8,12}$ 
I.~Valtchanov$^{6}$ 
\\
$^1$School of Physics \& Astronomy, Cardiff University, The Parade, Cardiff CF24 3AA, UK\\
$^2$California Institute of Technology, 1200 E.~California Blvd., Pasadena, CA 91125, USA\\
$^3$Infrared Processing and Analysis Center, MS 100-22, California Institute of Technology, JPL, Pasadena, CA 91125, USA\\ 
$^4$Jodrell Bank Centre for Astrophysics, University of Manchester, Alan Turing Building, Manchester M13 9PL, UK\\
$^5$Jet Propulsion Laboratory, 4800 Oak Grove Drive, Pasadena, CA 91109, USA\\
$^6$Herschel Science Centre, European Space Astronomy Centre, ESA, Villanueva de la Ca\~nada, E-28691 Madrid, Spain
$^7$Center for Astrophysics and Space Astronomy,  CB-389, University of Colorado, Boulder, CO 80309, USA\\
$^8$RALSpace, Science and Technology Facilities Council, Rutherford Appleton Laboratory,\\
~~Harwell Oxford, Didcot, Oxfordshire, OX11 0QX, UK\\
$^9$Department of Physical Science, The Open University, Milton Keynes, MK7 6AA, UK\\
$^{10}$Gemini Observatory Northern Operations Center, 670 N.~A'ohoku Place, Hilo, Hawaii, 96720, USA\\
$^{11}$SRON Netherlands Institute for Space Research, Landleven 12, 9747 AD Groningen, The Netherlands\\
$^{12}$Department of Physics \& Astronomy, University College London, Gower Place, London WC1E 6BT,UK\\
}

\begin{document}

\date{Accepted 2013. Received 2013.}

\pagerange{\pageref{firstpage}--\pageref{lastpage}} \pubyear{2013}

\maketitle

\label{firstpage}


\begin{abstract}
Photometric instruments operating at far infrared to millimetre
wavelengths often have broad spectral passbands
($\lambda/\Delta\lambda \sim 3$ or less), especially those operating
in space. A broad passband can result in significant variation of the
beam profile and aperture efficiency across the passband, effects
which thus far have not generally been taken into account in
the flux calibration of such instruments.  With absolute calibration
uncertainties associated with the brightness of primary calibration
standards now in the region of 5\% or less, variation of the
beam properties across the passband can be a significant contributor
to the overall calibration accuracy for extended emission. We present
a calibration framework which takes such variations into account for
both antenna-coupled and absorber-coupled focal plane
architectures. The scheme covers point source and extended source
cases, and also the intermediate case of a semi-extended source
profile. We apply the new method to the \spire\ space-borne
photometer.
\end{abstract}



\section{Introduction}
\label{sec:intro}
Broadband bolometric detector arrays are used in sensitive photometric
instruments for far infrared and submillimetre astronomy, particularly
in spaceborne instruments for which there are no atmospheric
limitations on the width of the accepted passband.  Considerations of
photometric sensitivity and optimum spectral definition lead to the
use of spectral passbands with resolution $\lambda/\Delta\lambda \sim$
2--3.  For such instruments, achieving the most accurate calibration
may require taking into account the variation of the beam profile and
aperture efficiency across the passband.  Recent and forthcoming
examples include the \spire\ \citep{Griffin2010_SPIRE}, \pacs\
\citep{Poglitsch2010_PACS} and \hfi\ \citep{Ade2010_HFIoptics}
satellite instruments, balloon-borne instruments such as BLAST
\citep{Pascale2008_BLAST}, PILOT \citep{Bernard2010_PILOT}, and EBEX
\citep{EBEX_SPIE2010}, and the proposed next-generation CMB
polarisation satellite mission, CORE \citep{COrE2011}.  Ground-based
instruments such as SCUBA \citep{Holland1999_SCUBA} and SCUBA-2
\citep{Holland2006_SCUBA2}, BOLOCAM \citep{Glenn2003_BOLOCAM}, LABOCA
\citep{Siringo2009_LABOCA}, SABOCA \citep{Siringo2010_SABOCA}, ACBAR
\citep{Runyan2003_ACBAR}, usually have narrower passbands
($\lambda/\Delta\lambda =$ 5--10) dictated by the widths of the
atmospheric windows in which they observe, although antenna-coupled
instruments operating at low frequency at high-transparency sites,
such as the South Pole Telescope, can have $\lambda/\Delta\lambda$ as
low as $\sim4$ \citep{Ruhl2004_SPT}.

Flux calibration in the far infrared (FIR) and submillimetre is
normally based on use of the planets Mars, Uranus or Neptune as the
primary standards (e.g., \citealt{Griffin1993_UranusNeptune}).  The
absolute uncertainty in planetary models is now at the level of ~5\%
(e.g., \citealt{MorenoThesis,Moreno2012}), and the stability and data
quality achievable, particularly with space-borne instruments, means
that relative calibration uncertainties can be less than this. These
improvements in the absolute knowledge of the primary standards and in
data quality mean that care needs to be taken in all aspects of flux
calibration to avoid systematic errors dominating the final
results. For broadband photometry, accurate knowledge of the
instrument relative Spectral Response Function (SRF) is essential, as
significant colour corrections must be applied to account for
differences in the spectral energy distributions of the calibrator and
the science target. The calibration of extended emission using a
point-like flux standard also requires good knowledge of the beam
properties. Close attention may also be needed to eliminate the
effects of non-linearity in the detector response. Normally the
conversion from point source to extended source calibration is
effected using an accurate estimation of the beam solid angle as
measured using a high signal-to-noise map of a bright point-like
source.  A complication, which has not yet been addressed, is that for
broadband detectors, the beam width varies significantly across the
passband.  The measured beam solid angle represents an average value
across the band, and also depends on the spectral shape of the object
used to map the beam. It is therefore not exactly the correct solid
angle to use when observing emission with a different spectral index
(which is usually the case).  Depending on the bandwidth and the
overall error budget, the variation of the beam profile across the
passband may or may not need to be taken into account in the
calibration procedure.

In this paper we outline an approach to flux calibration suitable for
point and extended sources, and use as an example the \spire\ camera,
which observes in three photometric bands centred near 250, 350, and
500\,\mic, and for which Neptune is used as the primary flux
standard. Detector non-linearity correction and many other practical
issues involved in making accurate comparative measurements of the
signal levels from the calibrator and the source are not the subject
of this paper and will not be further discussed here.  Such aspects
will be addressed for the case of SPIRE in \citet{Bendo2013} and Lim
et al. (in preparation).

In Section \ref{sec:ptsrc} the method adopted for point source
calibration is outlined and the formulae appropriate for the
derivation of a well-defined monochromatic flux density and for
appropriate colour correction are derived.  Section \ref{sec:extsrc}
outlines a scheme for the accurate calibration of extended or
semi-extended emission, which in principle requires knowledge of both
the beam properties of the system as a function of wavelength within
the spectral passband, and of the spatial distribution of the source
brightness. In Section \ref{sec:beamvar} we consider the manner in
which the aperture efficiency and beam profile vary across the
passband for both a single-moded feedhorn antenna-coupled detector and
an absorber-coupled detector, and in Section \ref{sec:resuniform} we
present results for the cases of ideal flat-topped passband of widths
$\lambda/\Delta\lambda =$3, 5 and 10. The application of this
methodology to \spire\ calibration is described and discussed in
Section \ref{sec:spire}. Conclusions are summarised in Section
\ref{sec:conc}, and the Appendix contains a list of symbols used in
the paper.


\section{Calibration of on-axis point source observations}
\label{sec:ptsrc}

Consider an instrument with a SRF defined by $\srf$, i.e., the
instrument transmission as a function of frequency, $\nu$ with
arbitrary normalisation.  Let $\eta(\nu)$ be the aperture efficiency,
defined as the fraction of the total power from an on-axis point
source that is coupled to the detector. Consider an on-axis
observation of a point source with spectral flux density $S(\nu)$ at
the telescope aperture.  The source power absorbed by the detector is
directly proportional to the integral over the passband of the flux
density weighted by the product of $\srf$ and $\eta(\nu)$.  Assuming
that the detector signal is, or can be linearised so as to be,
directly proportional to the absorbed power, the property of the
source that is measured, is the SRF-weighted flux density, given by

\begin{equation}
\label{eq:smeas_gen}
\smeas = \frac{{\displaystyle \intnu S(\nu) \srf \eta(\nu) \dd\nu}}
         {{\displaystyle \intnu \srf \eta(\nu) \dd\nu}} \quad.
\end{equation}

When observing a point-like calibrator, the quantity that is directly
proportional to absorbed detector power is the calibrator SRF-weighted
flux density, given by

\begin{equation}
\label{eq:scal}
\overline{S}_\mathrm{C} = K_\mathrm{Beam}(\theta_\mathrm{p},\theta_\mathrm{Beam}) \left[
  \frac{{\displaystyle \intnu S_\mathrm{C}(\nu) \srf \eta(\nu) \dd\nu}}
       {{\displaystyle \intnu \srf \eta(\nu) \dd\nu}}\right]\quad,
\end{equation}
where $S_\mathrm{C}(\nu)$ represents the calibrator spectrum and
$K_\mathrm{Beam}(\theta_\mathrm{p},\theta_\mathrm{Beam})$ is a
correction factor for possible partial resolution of the calibrator by
the telescope beam. For a Gaussian main beam profile coupling to a
uniformly bright disk (such as planet or asteroid) the beam correction
factor is given by \citet{Ulich1976_WaterCal}:

\begin{equation}
\label{eq:kbeam}
K_\mathrm{Beam}(\theta_\mathrm{p},\theta_\mathrm{Beam}) = 
\frac{{\displaystyle1 - \exp\biggl(-\frac{4\ln(2)\theta_\mathrm{p}^2}{\theta_\mathrm{Beam}^2}\biggr)}}
     {\displaystyle{\frac{4\ln(2)\theta_\mathrm{p}^2}{\theta_\mathrm{Beam}^2}}} \quad,
\end{equation}
where $\theta_\mathrm{p}$ is the angular radius of the disk, and
$\theta_\mathrm{Beam}$ is the beam full-width half-maximum value
(FWHM).  Because both the beam width and $\theta_\mathrm{p}$ can vary
across the passband, $K_\mathrm{Beam}$ can have a weak dependence on
frequency.

The calibration flux density, $S_\mathrm{C}$, can thus be derived from
a knowledge of the calibrator spectrum and angular size, as well as
the instrument SRF and aperture efficiency functions.  It is
convenient and conventional to quote the result of a photometric
measurement in the form of a monochromatic flux density at a suitable
standard frequency, $\nu_0$, near the centre of the passband.  If an
unknown source has the same spectral shape as the calibrator, then its
flux density at any frequency can be determined simply from the ratio
of the measured signals multiplied by the calibrator flux density at
that frequency.  In the more usual case in which the source and
calibrator have different spectral shapes, the definition of a
monochromatic flux density requires, in addition to a standard
frequency, some assumption about the shape of the source spectrum.

For an operating instrument, it is convenient to implement a standard
automatic pipeline which processes the observational data to produce
corresponding flux densities based on some default assumption
concerning the source spectrum:
\begin{equation}
  \label{eq:snu}
  S(\nu) = S(\nu_0) \cdot f(\nu,\nu_0) \quad,
\end{equation}
where $f(\nu, \nu_0)$ characterises the shape of the
spectrum. Commonly adopted assumptions for the spectral shape are a
power law with some spectral index, $\alpha$:
\begin{equation}
  \label{eq:fnu_alpha}
  f(\alpha,\nu,\nu_0) = \Bigl(\frac{\nu}{\nu_0}\Bigr)^\alpha \quad, 
\end{equation}
or a modified black body characterised by a black body spectrum
$\mathcal{B}(\nu,T)$ for temperature $T$, modified by an emissivity
that varies with frequency according to a power law with index
$\beta$:
\begin{equation}
  \label{eq:fnu_beta}
  \begin{split}
    f(T,\beta,\nu,\nu_0) &= \frac{\mathcal{B}(\nu,T)}{\mathcal{B}(\nu_0,T)} \biggl(\frac{\nu}{\nu_0}\biggr)^\beta \\
    &= \frac{{\displaystyle\exp\biggl(\frac{h\nu_0}{k_B T}\biggr) - 1}}
    {{\displaystyle\exp\biggl(\frac{h\nu}{k_B T}\biggr) - 1}}
    \biggl(\frac{\nu}{\nu_0}\biggr)^{3+\beta} \quad,
  \end{split}
\end{equation}
where $h$ is Planck's constant and $k_B$ is Boltzmann's constant. The
power law spectrum in \eqref{eq:fnu_alpha} is simple to estimate based
on empirical data from two or more photometric measurements, while the
modified black body spectrum in \eqref{eq:fnu_beta} is a simple model
representing optically thin thermal dust emission.

From \eqref{eq:smeas_gen} and \eqref{eq:snu}, and assuming the case of
a source with a power-law spectral energy distribution (SED) given by
$f(\alpha,\nu,\nu_0)$ as in \eqref{eq:fnu_alpha}, the monochromatic
source flux density at frequency $\nu_0$ is given by
\begin{equation}
  \label{eq:s_kmonp}
  S(\nu_0) = \kmonp(f,\nu_0) \cdot \smeas \quad,
\end{equation}
with
\begin{equation}
  \label{eq:kmonp}
  \kmonp(\alpha,\nu_0) = 
  \frac{{\displaystyle \intnu \srf \eta(\nu) \dd\nu}}
  {{\displaystyle \intnu \biggl(\frac{\nu}{\nu_0}\biggr)^\alpha \srf \eta(\nu) \dd\nu}} \quad,
\end{equation}
while for a modified black body spectrum, $\kmonp$ is a function of
$(T,\beta,\nu_0)$ insead of $(\alpha,\nu_0)$.

To derive the monochromatic flux density at frequency $\nu_0$, the
measured SRF-weighted flux density of a point source is therefore
multiplied by $\kmonp$, which can be computed from the known
properties of the instrument and an assumed source spectral shape.  In
pipeline processing of instrument data, a power law spectrum with a
standard value of $\alpha_0=-1$ is often adopted, corresponding to a
source spectrum which is flat in $\nu S(\nu)$, i.e., with
frequency-independent power flux.  This is the convention used in the
generation of pipeline flux densities for many previous instruments in
the mid-infrared through millimetre, including {\em COBE}/DIRBE
\citep{COBE_ExpSupp}, {\em IRAS} \citep{Beichman1988_IRAS}, ISOCAM
\citep{cesarsky1996_ISOCAM}, ISOPHOT \citep{lemke1996_ISOPHOT}, and
for the \herschel\ photometers PACS \citep{Poglitsch2010_PACS} and
SPIRE \citep{Swinyard2010_SPIRECal}. An exception is \spitzer-MIPS,
which uses a $10^4$\,K black body as a reference spectrum
\citep{Stansberry2007_MIPScal}.  This assumption results in a pipeline
flux density
\begin{equation}
  \label{eq:spip}
  \spip(\alpha_0,\nu_0) = 
  \kmonp(\alpha_0,\nu_0) \cdot \smeas \quad.
\end{equation}

The assumption that the source has a power law spectrum with a
spectral index $\alpha_0$ will not be valid in most cases, requiring
the application of a point source colour correction factor,
$\kcolp$, based on the best available information on the
actual source spectrum (for instance, measurements at multiple
wavelengths).  The true monochromatic flux density is then estimated
as
\begin{equation}
  \label{eq:snu0_kcolp}
  S(\nu_0) = \kcolp(f,\alpha_0,\nu_0) \cdot \spip(\nu_0) \quad.
\end{equation}

For an assumed power-law spectrum, with $f(\alpha,\nu,\nu_0)$ as given
in \eqref{eq:fnu_alpha},
\begin{equation}
  \label{eq:kcolp_a}
  \begin{split}
    \kcolp(\alpha,\alpha_0,\nu_0) &= 
    \frac{{\displaystyle \kmonp(\alpha,\nu_0)}}
      {{\displaystyle \kmonp(\alpha_0,\nu_0)}} \\
    &= \left[
    \frac{{\displaystyle \intnu \srf \eta(\nu) \nu^{\alpha_0} \dd\nu}}
      {{\displaystyle \intnu \srf \eta(\nu) \nu^\alpha \dd\nu}}\right]
    \nu_0^{\alpha - \alpha_0} \quad.
  \end{split}
\end{equation}

In the case of an assumed modified black body spectrum, with
$f(T,\beta,\nu,\nu_0)$ as given in \eqref{eq:fnu_beta},
\begin{multline}
  \label{eq:kcolp_b}
  \kcolp(T,\beta,\alpha_0,\nu_0) = 
  \frac{{\displaystyle \nu_0^{3 + \beta - \alpha_0}}}
       {{\displaystyle e^{h \nu_0/k_B T} - 1}}\\
       \times
       \left[ \frac{{\displaystyle \intnu \nu^{\alpha_0} \srf \eta(\nu) \dd\nu}}
         {{\displaystyle \intnu \biggl(
             \frac{\nu^{3 + \beta}}{e^{h \nu/k_B T} - 1} \biggr)
             \srf \eta(\nu) \dd\nu}}
         \right] \quad.
\end{multline}

If $h\nu \ll k_B T$ (i.e., a modified black body in the Rayleigh-Jeans regime),
this is equivalent to the power law case with $\alpha=\beta + 2$.

\section{Calibration of extended emission}
\label{sec:extsrc}

A conventional approach to the calibration of extended emission is to
divide the measured flux densities (based on a point source
calibration scheme) by the solid angle of the beam to convert a map
from units of flux density (Jy in beam) to units of surface brightness
(e.g.~Jy/pixel or Jy/sr).  This procedure is potentially inaccurate if
the beam profile varies as a function of frequency within the band, in
which case it is not valid to define a particular beam area. As
discussed in Section 4 below, in the case of broadband photometric
observations the beam can vary significantly across the passband.  To
account correctly for the wavelength dependence of the beam, it is
then necessary to include it explicitly in the integral for the
measured flux density.

Consider a point source observation in which the source is not on
axis, but at some angular position with respect to the axis, defined
by orthogonal offset angles $\theta$ and $\phi$. Let the
frequency-dependent normalised beam response as a function of position
be $B(\nu,\theta,\phi)$, so that at a given frequency, $\nu$, the
response to the source at position $(\theta,\phi)$ is reduced by that
factor with respect to the on-axis value at that frequency.
A slightly extended source at $(\theta,\phi)$, with surface brightness
$I(\nu,\theta,\phi)$ and angular extent $(\dd\theta,\dd\phi)$, will
therefore lead to a measured flux density given by
\begin{equation}
  \label{eq:dsmeas_ext}
  \dd\smeas(\theta,\phi) = 
  \frac{{\displaystyle \intnu I(\nu,\theta,\phi) B(\nu,\theta,\phi) \dd\theta \dd\phi \,
      \srf \eta(\nu) \dd\nu}}
       {{\displaystyle \intnu \srf \eta(\nu) \dd\nu}} \quad.
\end{equation}

The total measured flux density within the entire beam for an extended
source is then:
\begin{equation}
  \label{eq:smeas_ext}
  \smeas =
  \frac{{\displaystyle \intnu y(\nu) \srf \eta(\nu) \dd\nu}}
       {{\displaystyle \intnu \srf \eta(\nu) \dd\nu}} \quad.
\end{equation}
where
\begin{equation}
  \label{eq:y}
  y(\nu) = \intbeam I(\nu,\theta,\phi) B(\nu,\theta,\phi) \dd\theta \dd\phi
\end{equation}

To derive an estimate of a monochromatic sky surface brightness from
the broadband measurement, it is therefore necessary (i) to have
knowledge of the beam response as a function of frequency
across the band, (ii) to assume some particular spectral shape
for the source emission, and (iii) to assume some particular source
spatial distribution.  For simplicity, we consider here the case in
which both the beam and the source are circularly symmetric, and the
source surface brightness profile $I(\nu,\theta)$ varies with position
according to some function $g(\theta,\theta_0)$ relative to a scale
radius $\theta_0$, and with frequency according to $f(\nu,\nu_0)$, but
has the same frequency dependence for all positions:
\begin{equation}
  \label{eq:inutheta}
  I(\nu,\theta) = I(\nu_0,0) \cdot f(\nu,\nu_0) \cdot g(\theta,\theta_0) \quad.
\end{equation}

As before a power law or a modified black body could be used for $f(\nu,\nu_0)$.
A reasonable assumption for the spatial variation could be another
power law or a Gaussian function.

The peak surface brightness at frequency $\nu_0$ is then
\begin{equation}
  \label{eq:inu0}
  I(\nu_0,0) = \kmone(f,g,\nu_0) \cdot \smeas \quad,
\end{equation}
where
\begin{equation}
  \label{eq:kmone}
  \kmone(f,g,\nu_0) =
  \frac{{\displaystyle \intnu \srf \eta(\nu) \dd\nu}}
       {{\displaystyle \intnu y^\prime(\nu,\theta_0) f(\nu,\nu_0) \srf \eta(\nu) \dd\nu}} \quad,
\end{equation}
with
\begin{equation}
  \label{eq:yprime}
  y^\prime(\nu,\theta_0) = \intbeam P(\nu,\theta) g(\theta,\theta_0) 2 \pi \theta \dd\theta \quad,
\end{equation}
where $P(\nu,\theta)$ is the circularly-symmetic beam profile.

It is important to note that, unlike the point source case, $\kmone$
is not dimensionless, but converts from flux density (Jy, or Jy/beam
for extended sources) to surface brightness (conventionally MJy/sr).

An alternative to the point source calibration scheme described in the
previous section would be to assume, as standard, fully extended
emission and to derive the sky surface brightness using
\eqref{eq:inu0}. Note that in the case of uniform extended emission
filling the beam, where $g(\theta,\theta_0)=1$, the integral over the
beam reduces to the frequency-dependent beam solid angle,
$\Omega(\nu)$, so that
\begin{multline}
  \label{eq:kuni}
  \kuni(f,\nu_0) \equiv \kmone(f,g=1,\nu_0) = \\
  \frac{{\displaystyle \intnu \srf \eta(\nu) \dd\nu}}
       {\Omega(\nu_0){\displaystyle \intnu \onorm(\nu,\nu_0) f(\nu,\nu_0)
           \srf \eta(\nu) \dd\nu}} \quad,
\end{multline}
where $\onorm(\nu,\nu_0)=\Omega(\nu)/\Omega(\nu_0)$ is the beam solid angle
normalised to the value at $\nu_0$. It is also possible to construct
an ``effective beam solid angle'' for a given source spectrum,
$\oeff(f)$:
\begin{equation}
  \label{eq:oeff}
  \oeff(f) = \frac{\displaystyle \intnu f(\nu,\nu_0) \Omega(\nu) \srf \eta(\nu) \dd\nu}
    {\displaystyle \intnu \srf \eta(\nu) \dd\nu} \quad.
\end{equation}

It may be the case that different processing steps are required for
point sources and extended emission. Automatic instrument pipelines
must therefore adopt one convention or the other, requiring a
post-pipeline correction in the case of a different source type.  For
example, if the pipeline processing for an extended source is based on
the assumption that the observation is of fully extended emission
(i.e.~$\theta_0=\infty$) which has a power-law spectrum with
$\alpha=\alpha_0$, then the extended source pipeline produces a map in
which the peak surface brightness is:
\begin{equation}
  \label{eq:ipip}
  \ipip(\alpha_0,\nu_0,\theta=0) = \kuni(\alpha_0,\nu_0) \cdot \smeas(\theta=0)
\end{equation}

The conversion from the point source to fully-extended source pipelines
is relatively straightforward:
\begin{equation}
  \label{eq:spip_ptoe}
  \ipip(\alpha_0,\nu_0,\theta=0) = \left[\frac{\kuni(\alpha_0,\nu_0)}{\kmonp(\alpha_0,\nu_0)}\right] \spip \quad.
\end{equation}

The peak surface brightness at frequency $\nu_0$ for a general
extended source can then be derived from the peak of the
extended pipeline surface brightness as
\begin{equation}
  \label{eq:inu0kext}
  I(\nu_0,0) = \kcole(f,g,\alpha_0,\nu_0) \cdot \ipip(\alpha_0,\nu_0,\theta=0) \quad,
\end{equation}
where
\begin{equation}
  \label{eq:kcole}
  \begin{array}{l}
    \kcole(f,g,\alpha_0,\nu_0) =\frac{{\displaystyle \kmone(f,g,\nu_0)}}{{\displaystyle \kuni(\alpha_0,\nu_0)}} \\ \\
    \quad= \frac{{\displaystyle \Omega(\nu_0) \intnu \onorm(\nu,\nu_0) \biggl(\frac{\nu}{\nu_0}\biggr)^{\alpha_0} \srf \eta(\nu) \dd\nu}}
            {\displaystyle \intnu y^\prime(\nu,\theta_0) f(\nu,\nu_0) \srf \eta(\nu) \dd\nu} \quad.
  \end{array}
\end{equation}
where $y^\prime(\nu,\theta_0)$ is as defined in \eqref{eq:yprime}. The
variable $\kcole$ converts the extended pipeline surface brightness,
which assumes a fully extended source with spectral index $\alpha_0$,
to that for an extended source with brightness profile
$g(\theta,\theta_0)$ and spectrum $f(\nu,\nu_0)$.

The monochromatic beam profile and beam solid angle are not normally
measured directly, but have to be modelled based on the instrument
design and broadband measurements of a particular source. The
broadband beam profile is usually determined by mapping a point-like
source such as a planet.  If the source used for the beam mapping has
spectral index $\alpha_p$, then the resulting measured beam profile is
\begin{equation}
  \label{eq:pmeas}
  \pmeas(\theta,\alpha_p) = 
  \frac{{\displaystyle \intnu \nu^{\alpha_p} P(\nu,\theta) \srf \eta(\nu) \dd\nu}}
  {{\displaystyle \intnu \nu^{\alpha_p} \srf \eta(\nu) \dd\nu}} \quad,
\end{equation}
and the corresponding measured beam solid angle is
\begin{equation}
  \label{eq:omeas}
  \omeas(\alpha_p) = \intbeam \pmeas(\theta,\alpha_p) \, 2\pi\theta\dd\theta \quad.
\end{equation}

This method assumes a point-like source, so it may also be necessary
to apply a correction for the finite angular size of the calibration
target.

\section{Beam profile and aperture efficiency variation within the passband}
\label{sec:beamvar}
The ways in which the beam profile and the aperture efficiency vary
across the passband depend on the detector array architecture, in
particular on whether the detectors are antenna-coupled (single-moded)
or absorber-coupled (multi-moded).  Here we consider two of the most
commonly adopted configurations: a feedhorn antenna-coupled system
designed to have a horn aperture corresponding to $2\ld$ (where $D$ is
the telescope aperture) at the band centre in order to give maximum
aperture efficiency, and an absorber-coupled system designed to have
square pixels of side $0.5\ld$ at the band centre in order to provide
instantaneous Nyquist sampling of the sky image.  In both cases we
consider an idealised passband of width given by $R = \nu/\Delta\nu =
\lambda/\Delta\lambda = 3$, with constant transmission within the band
and zero outside.

\subsection{Antenna-coupled case}
\label{sec:antenna}

For a diffraction-limited antenna-coupled system (e.g. \spire, BLAST),
the beam FWHM increases with increasing wavelength within the band due
to diffraction at the primary aperture (which gets smaller in relation
to the wavelength).  In the case of a wavelength-independent
illumination of the primary aperture by the detector, this variation
would be linear. However, for antenna coupling, it is necessary to
take into account the fact that the beam profile of the feed antenna
itself broadens with increasing wavelength, so that the aperture
illumination profile tends to be less tapered at longer wavelengths.
This means that the outer parts of the antenna are more strongly
illuminated at longer wavelengths, tending to make the beam narrower
and offsetting the effect of diffraction at the primary.  The expected
variation of beam FWHM with wavelength is therefore slower than
linear.  To examine this effect, we consider the idealised case of a
feed antenna illuminating an unobscured telescope with a Gaussian
illumination pattern of width directly proportional to wavelength.
The edge taper (relative illumination at the primary edge compared to
the on-axis value) is taken to be 8\,dB at the centre of a band of
width $R = \nu/\Delta\nu = 3$.

\begin{figure}
  \centering
  \includegraphics[width=0.45\textwidth]{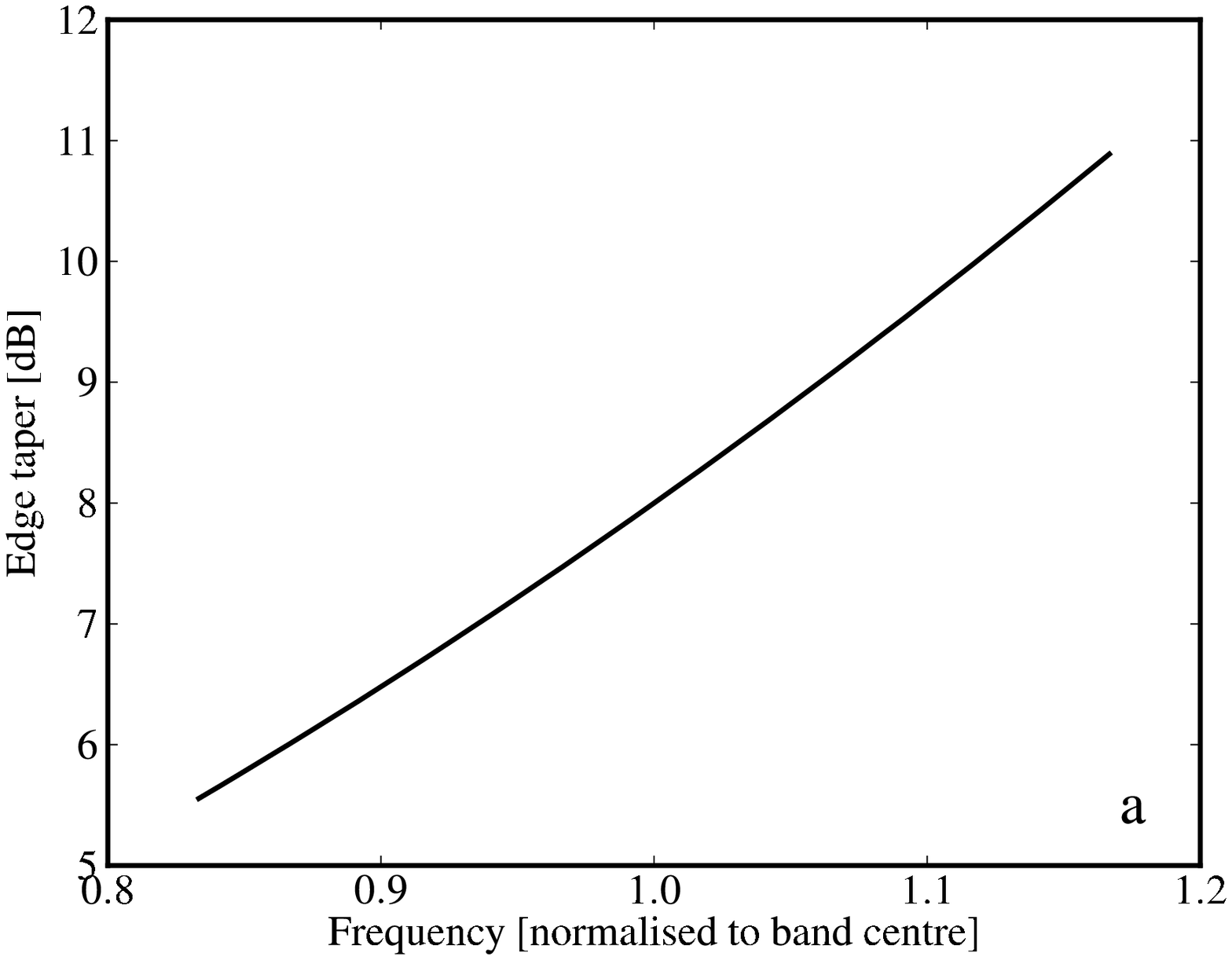}
  \includegraphics[width=0.45\textwidth]{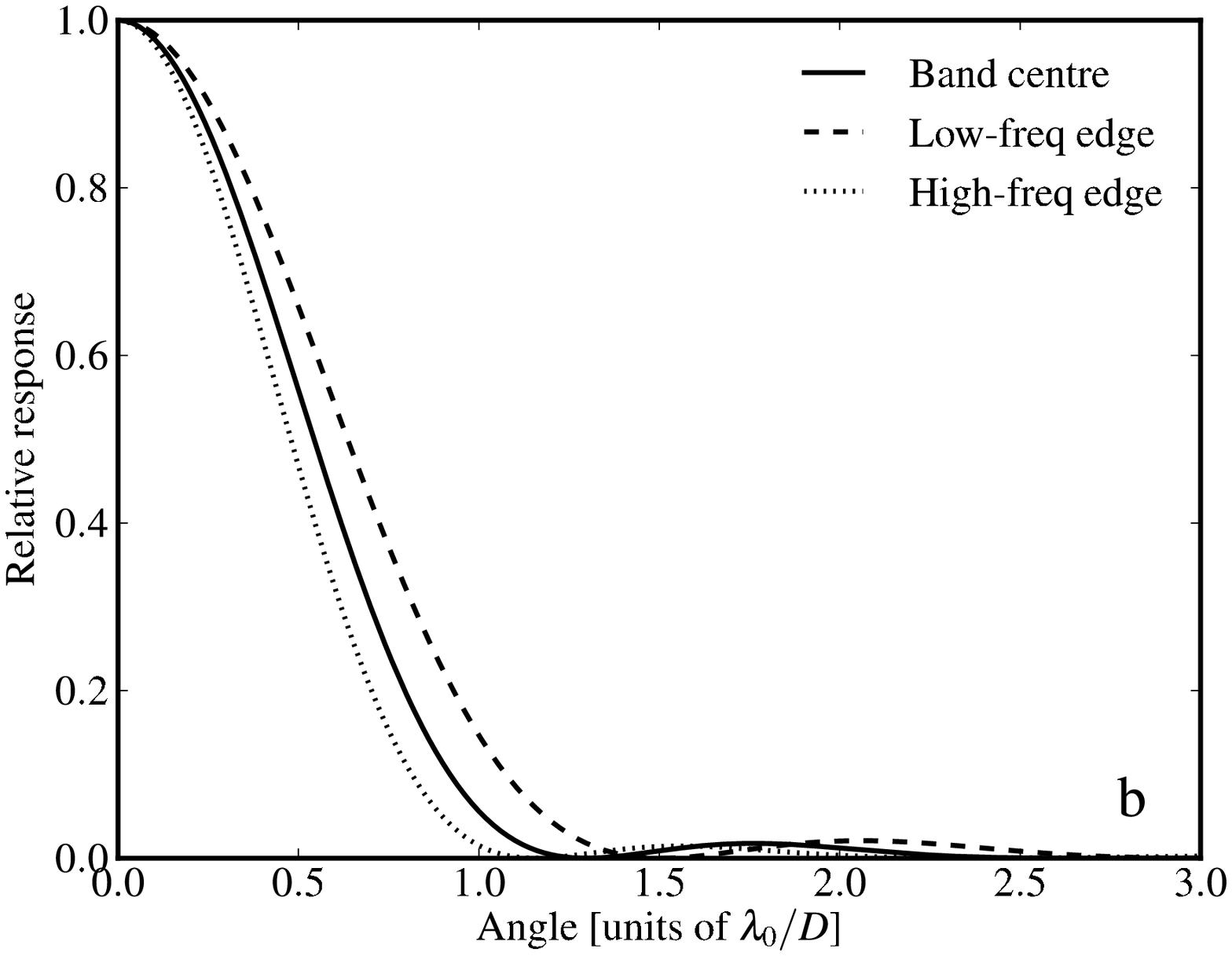}
  \caption{(a) Edge taper of the primary illumination vs. normalised
    frequency for the case of a feedhorn producing a Gaussian beam of
    width inversely proportional to frequency.  An edge taper of 8\,dB
    is adopted at the central wavelength of the band, which is taken
    to have $R=3$.  (b) Corresponding beam profiles at the band centre
    (solid line), low-frequency edge (dashed line) and high-frequency
    edge (dotted line).}
  \label{fig:edge-taper_response}
\end{figure}

\begin{figure}
  \centering
  \includegraphics[width=0.45\textwidth]{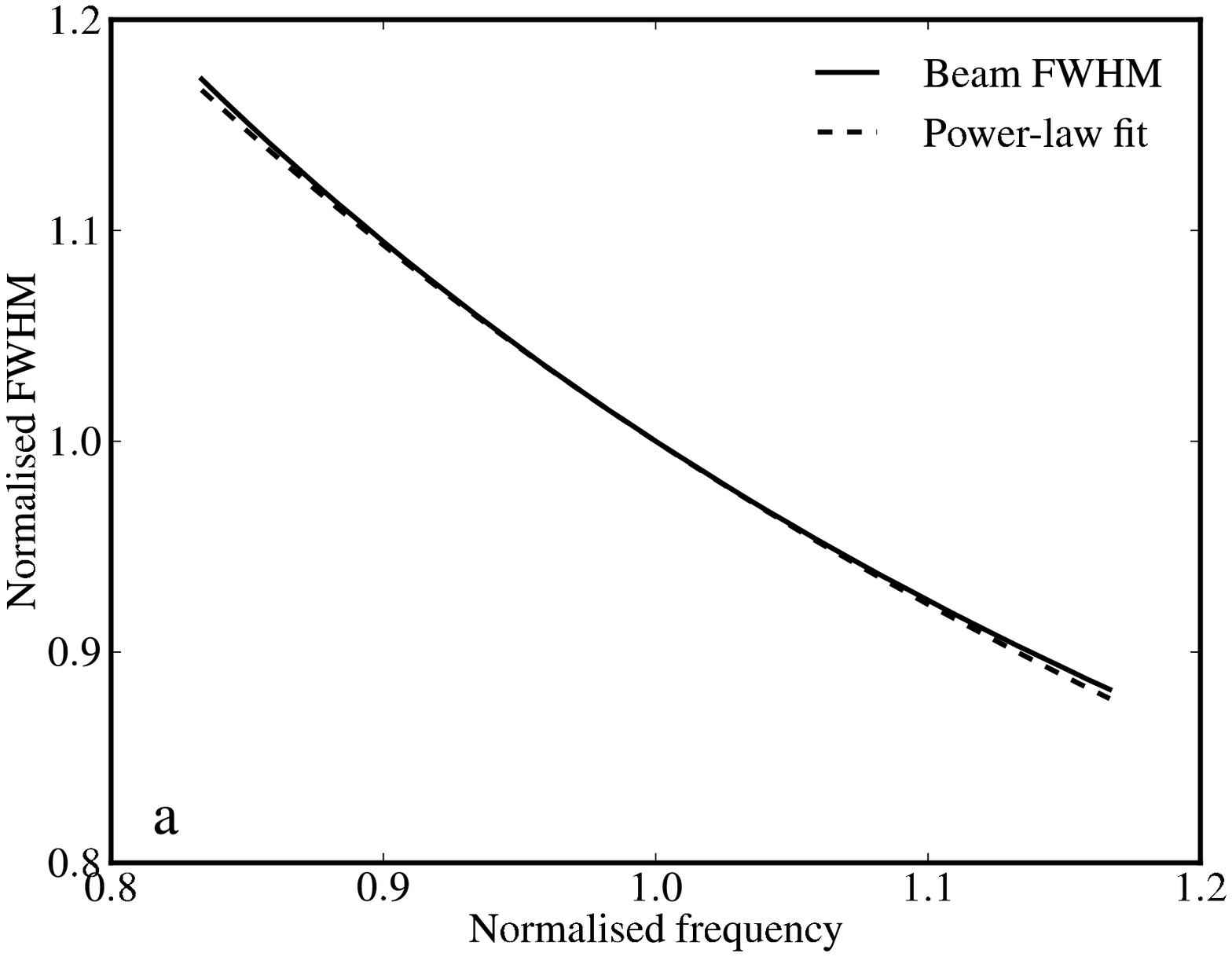}
  \includegraphics[width=0.45\textwidth]{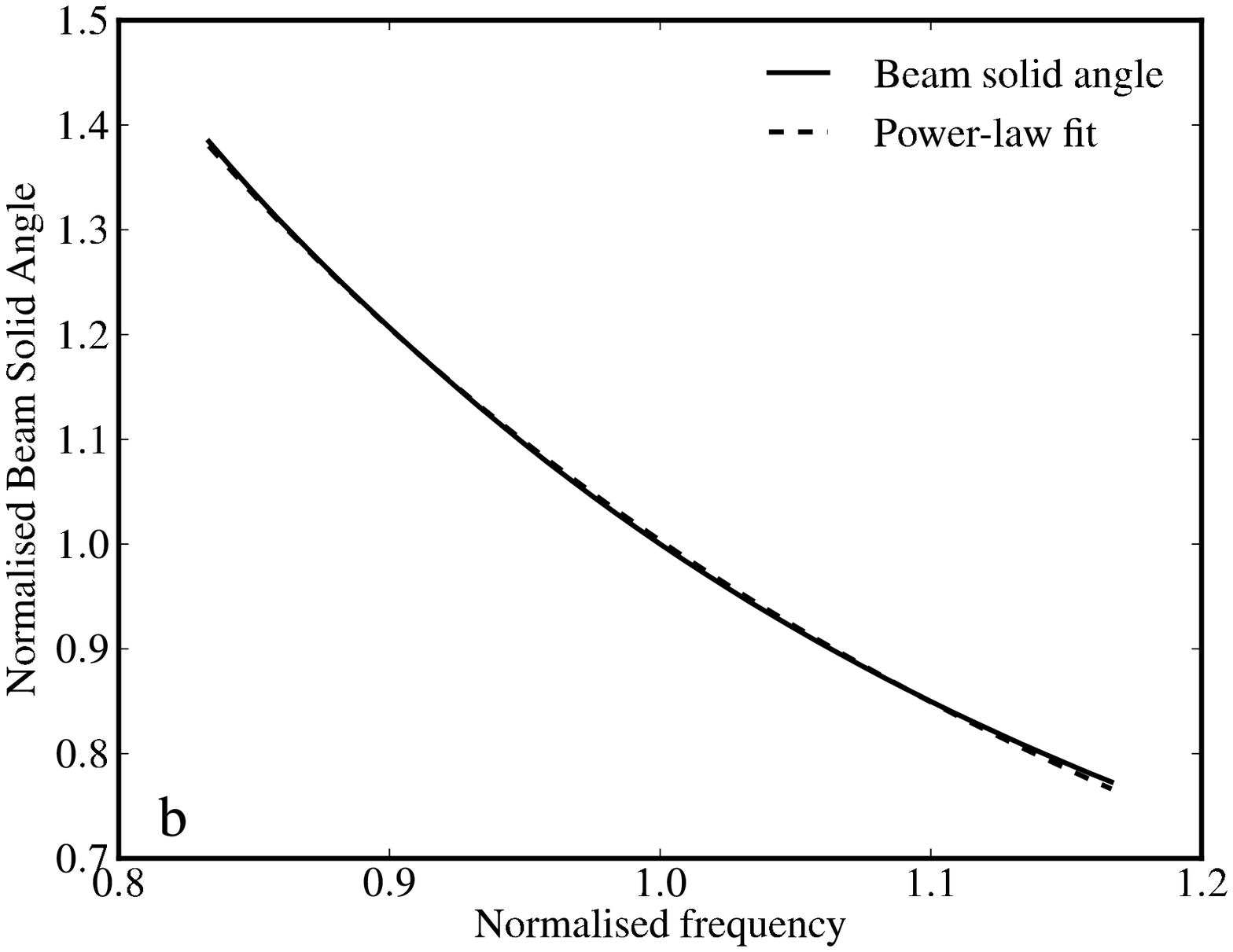}
  \caption{(a) Solid line: far-field beam FWHM (normalised to the
    value at the centre of the band) vs.~frequency (also normalised to
    the band centre) for Gaussian beam illumination of the primary
    with 8\,dB edge taper at the band centre.  Dashed line: power law
    fit where $\mathrm{FWHM}\propto\nu^\gamma$, with $\gamma=-0.85$.
    (b) Solid line: beam solid angle (normalised to the value at the
    centre of the band) vs.~frequency (also normalised to the band
    centre) for Gaussian beam illumination of the primary with 8\,dB
    edge taper at the band centre.  Dashed line: power law fit where
    $\Omega \propto \nu^\delta$, with $\delta=-1.75$.}
  \label{fig:fwhm_area}
\end{figure}

\figref{fig:edge-taper_response}(a) shows the edge taper of the
primary illumination vs.~normalised frequency.  The edge taper varies
from 5.9\,dB at the low-frequency edge of the band to 11.5\,dB at the
high-frequency end.  The telescope far-field beam profile is
calculated as the Fourier transform of the telescope illumination
pattern and is shown in \figref{fig:edge-taper_response}(b) for the
band centre and edges (with the angular offset in units of $\lod$,
where $\lambda_0$ is the band centre). The beam FWHM is plotted
in \figref{fig:fwhm_area}(a) as a function of frequency within the
band.  The variation is well fitted (residuals $<0.5\%$ in any part of
the band) by a power law: $\mathrm{FWHM} \propto \nu^\gamma$ with
$\gamma = -0.85$.  The FWHM varies by around $\pm17\%$ across the
passband.  The normalised beam solid angle, $\onorm$, is plotted
vs.~frequency in \figref{fig:fwhm_area}(b).  This variation is well
fitted (residuals $<1\%$ in any part of the band) by a power law
($\onorm \propto \nu^\delta$) with $\delta= 1.75$.

\begin{figure}
  \centering
  \includegraphics[width=0.45\textwidth]{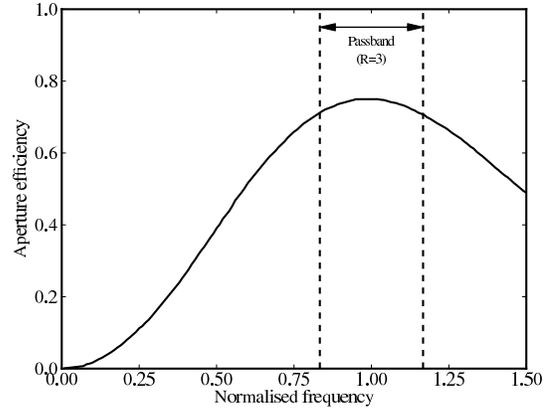}
  \caption{Aperture efficiency vs.~normalised frequency for a conical
    feedhorn antenna sized to have an entrance aperture of $2\ld$ at
    the centre of the passband.  The vertical dashed lines indicate a
    passband with $R = 3$.}
  \label{fig:apeff_ant}
\end{figure}

The on-axis aperture efficiency, $\eta$, for the case of a
smooth-walled conical horn is given in \citet{Griffin2002_FParch} as a
function of horn aperture diameter in units of $\ld$.  For a given
wavelength, aperture efficiency increases with aperture diameter,
reaches a broad maximum of $\sim0.75$ at approximately $2\ld$, and
declines for larger aperture size as the feedhorn beam narrows and
significantly under-illuminates the telescope resulting in a
diminution of efficiency.  The corresponding aperture efficiency as a
function of frequency for a feedhorn which is $2\ld$ at the centre of
a band with $R=3$ is shown in \figref{fig:apeff_ant}.  The aperture
efficiency has a broad peak and is fairly uniform across the band.

\subsection{Absorber-coupled case}
\label{sec:absorber}

\begin{figure}
  \centering
  \includegraphics[width=0.45\textwidth]{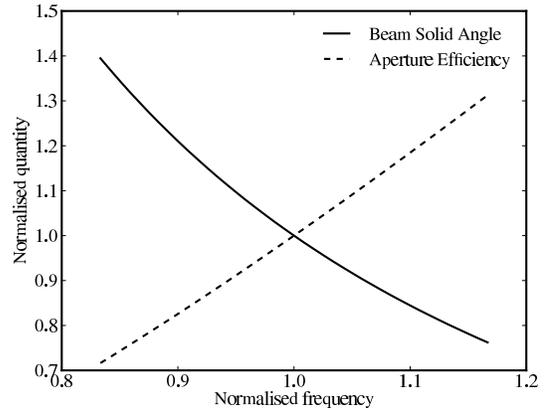}
  \caption{Beam solid angle (solid line), aperture efficiency (dashed
    line) vs.~frequency within an $R = 3$ passband for a pixel side of
    $0.5\ld$ at the band centre.  Both quantities are normalised to
    those at the band centre, at which frequency the aperture
    efficiency is $0.178$.}
  \label{fig:abs_area_apeff}
\end{figure}

The beam solid angle and aperture efficiency in the case of an
absorber-coupled detector (e.g.~\pacs, SCUBA2) have been computed by
convolving the Airy function for diffraction at a circular telescope
aperture with a square pixel of side $0.5\ld$ at the band centre.  The
results are shown in \figref{fig:abs_area_apeff}.  Although the
aperture efficiency and beam solid angle are both strongly
frequency-dependent, the product of the two, which determines how the
detector couples to fully extended emission, is constant across the
band (which must be the case in order to satisfy the requirement that
for sky intensity independent of position or frequency, the power per
unit frequency intercepted by a pixel must be constant across the
band).  This is in contrast to the antenna-coupled case described
above, where the aperture efficiency does not change much across the
band but the beam solid angle is strongly frequency-dependent, leading
to a much greater variation of coupling efficiency to extended
emission over the passband.  In the case of fully extended emission
with an absorber-coupled detector, it is not necessary to know the
frequency dependences of the beam solid angle and aperture efficiency
independently.  However, for semi-extended emission they must both be
taken into account explicitly.

\section{Example: calibration of a uniform passband}
\label{sec:resuniform}
As an example we take the case of a passband with $R = 3$, an assumed
source spectral index $\alpha_0=-1$, and nominal fequency $\nu_0$ at
the band centre, and either a point or fully extended source, and
consider the corresponding correction factors as a function of assumed
power-law source spectral index, for both the feedhorn and
absorber-coupled cases.

\begin{figure}
  \centering
  \includegraphics[width=0.45\textwidth]{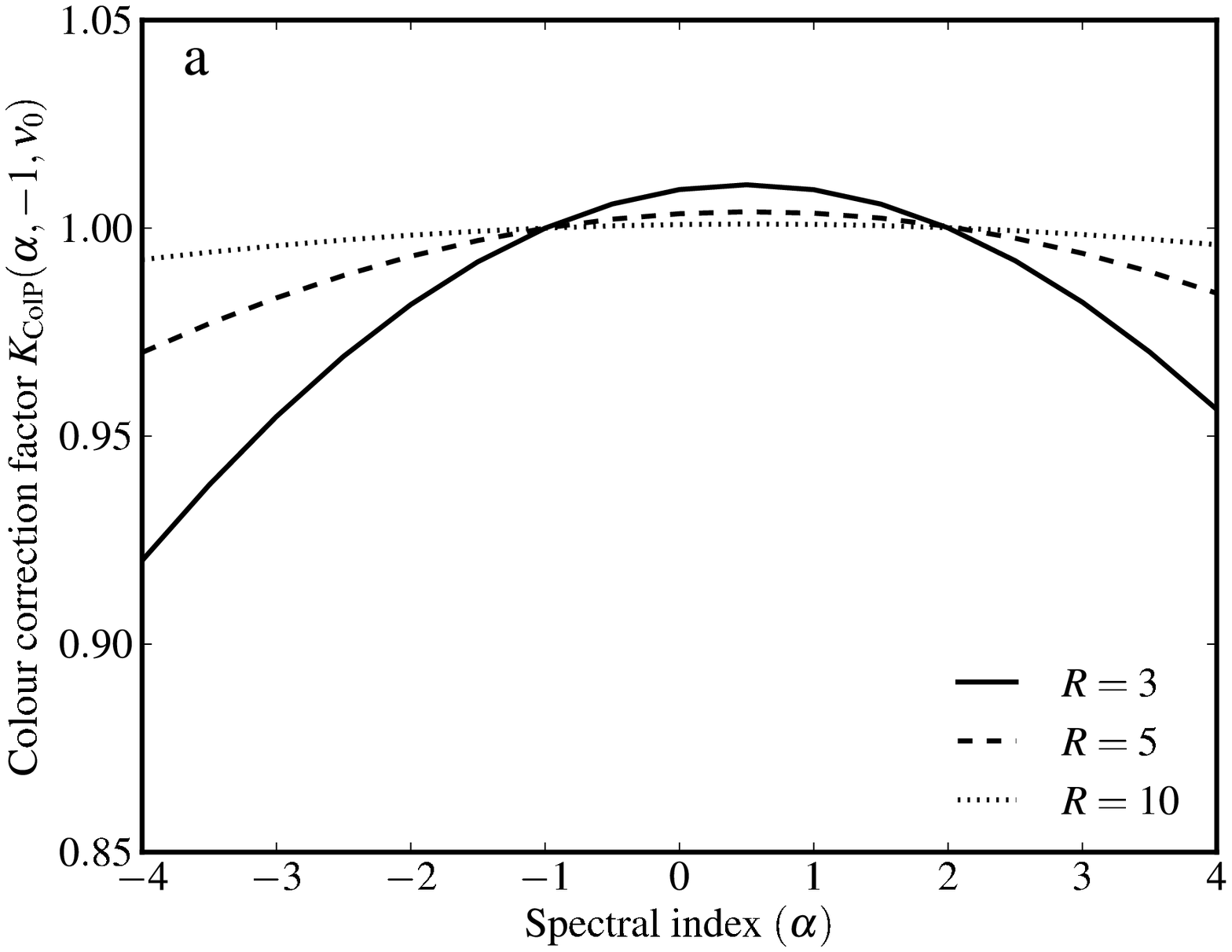}  
  \includegraphics[width=0.45\textwidth]{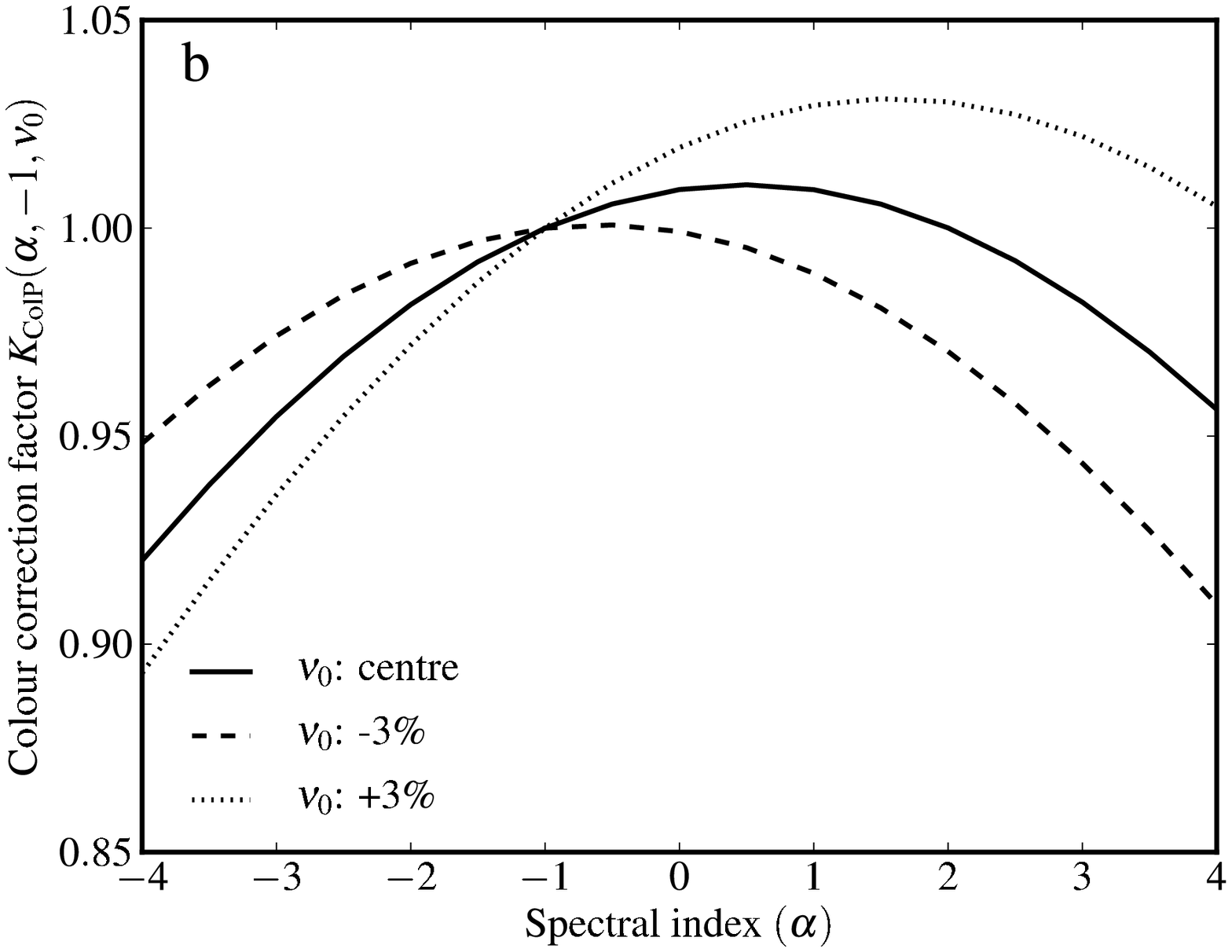}
  \includegraphics[width=0.45\textwidth]{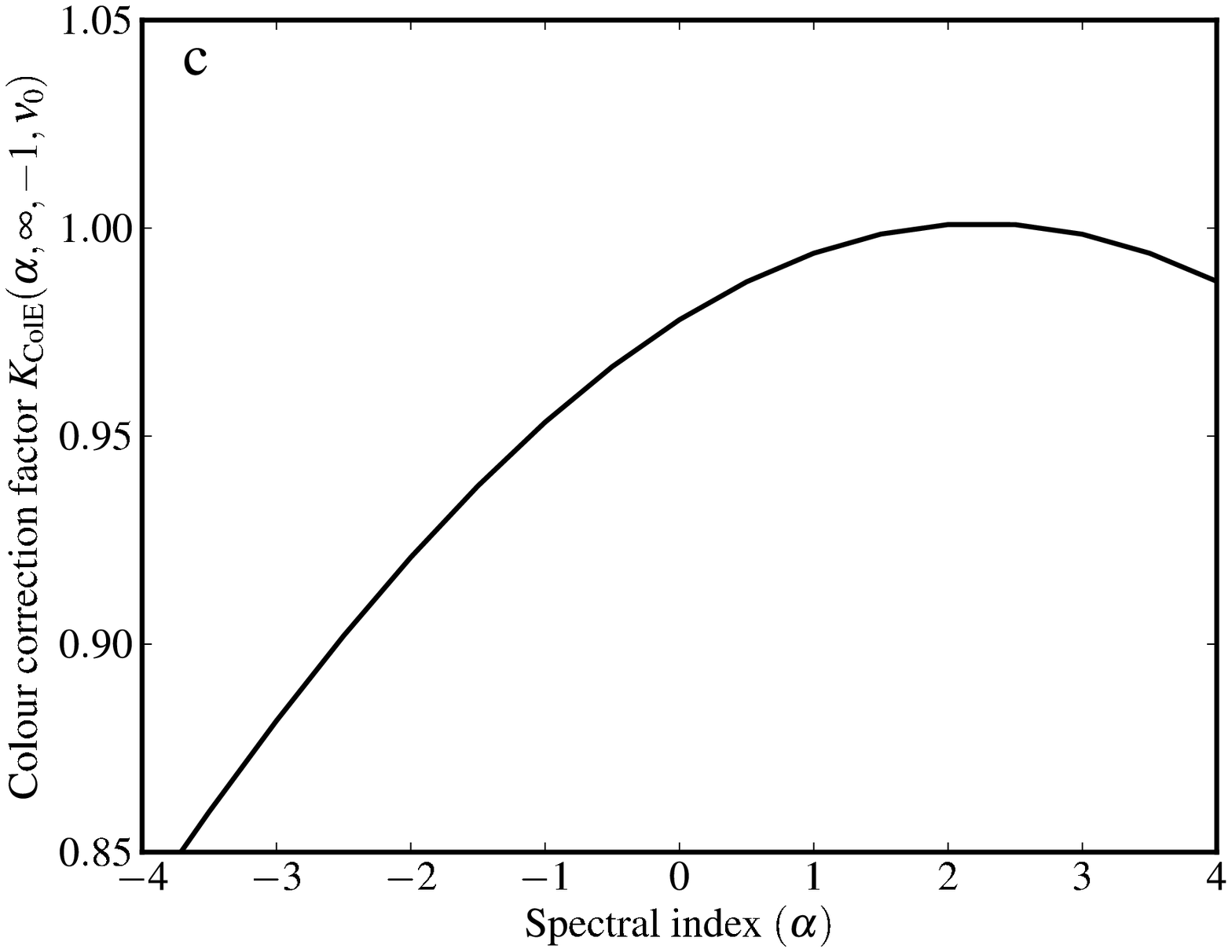}
  \caption{(a)~Point source colour correction factor vs.~power-law
    source spectral index for a conical feedhorn-coupled system with
    $R = 3$ (solid line), 5 (dashed line) and 10 (dotted line) and
    nominal frequency at the band centre. (b)~The same with $R = 3$
    and nominal frequency at the band centre (solid line), shifted by
    3\% of the bandwidth towards the low-frequency edge (dashed line),
    and 3\% towards the high-frequency edge (dotted
    line). (c)~Extended colour-correction conversion factor (conical
    feedhorn case; $R = 3$; nominal frequency at band centre; fully
    extended source) vs. assumed source spectral index.}
  \label{fig:ant_uni}
\end{figure}

The point source colour correction factor for an assumed power law
spectrum, $\kcolp(\alpha,-1,\nu_0)$, as computed using
\eqref{eq:kcolp_a}, is shown as a function of source spectral index in
\figref{fig:ant_uni}a for the antenna-coupled case.  Also shown in
\figref{fig:ant_uni}a are the colour correction curves for narrower
passbands with $R=5$ and $R=10$. The correction factor is seen to be
sensitive to the width of the passband. Moving the frequency of the
band edges by 1\% results in a 1.6\% change in the colour correction
factor for a source with spectral index $\alpha=3$.  It is found to be
much less sensitive to the shape of the passband or the aperture
efficiency function: changing the flat passband shape to one with a
$\pm$5\% tilt across the band results in only a 0.5\% change in the
colour correction factor. This insensitivity arises from the fact that
the SRF integral appears in both the numerator and denominator of
e.g.~\eqref{eq:kmonp}. The choice of the standard frequency, $\nu_0$,
also affects the correction significantly, as shown in
\figref{fig:ant_uni}b which illustrates the effect of shifting $\nu_0$
in either direction by 3\% of the bandwidth.

The colour correction factor for fully extended sources,
$\kcole(\alpha,\infty,-1,\nu_0)$, as given by \eqref{eq:kcole}, for
the feedhorn case with an $R=3$ passband, is plotted vs. assumed
source spectral index in \figref{fig:ant_uni}c.  Depending on the
source spectral index and the choice of nominal frequency, corrections
on the order of 10\% may be needed.  The correction factors are not
very sensitive to the exact value assumed for $\delta$, the power-law
for the solid angle variation across the band.  Changing $\delta$ by
$\pm10\%$ changes the correction factor by less than 1\% over the
whole range of the plot.

%
%

\begin{figure}
  \centering
  \includegraphics[width=0.45\textwidth]{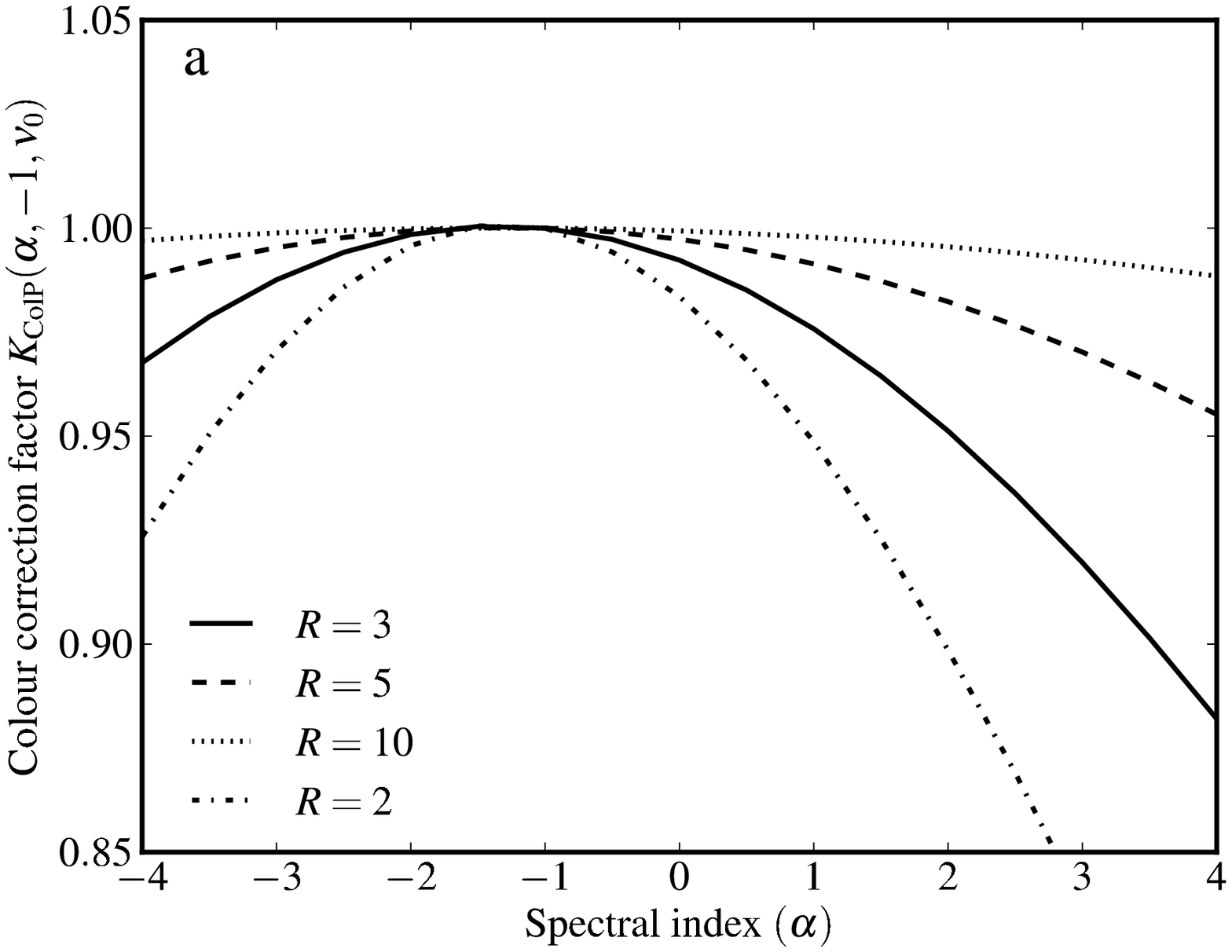}
  \includegraphics[width=0.45\textwidth]{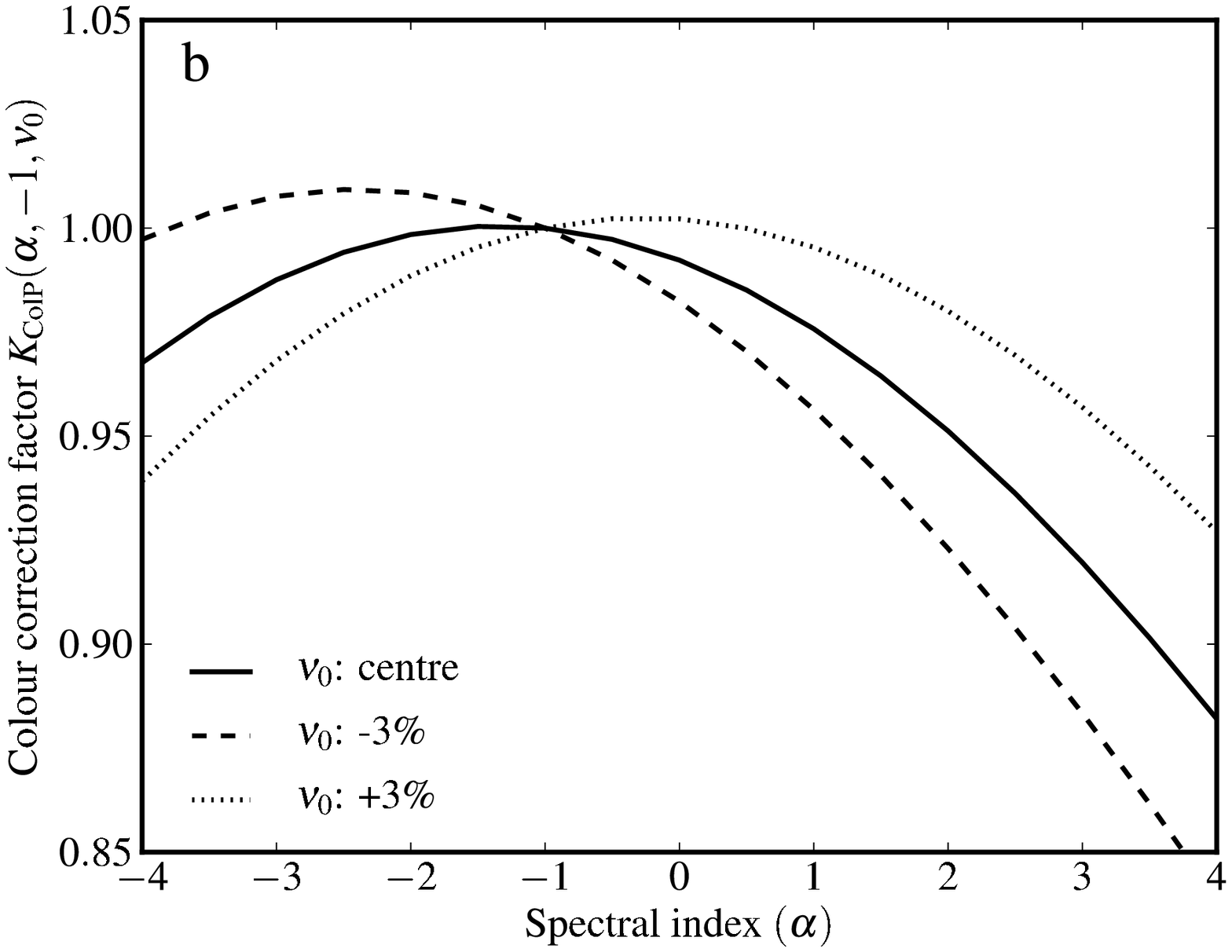}
  \includegraphics[width=0.45\textwidth]{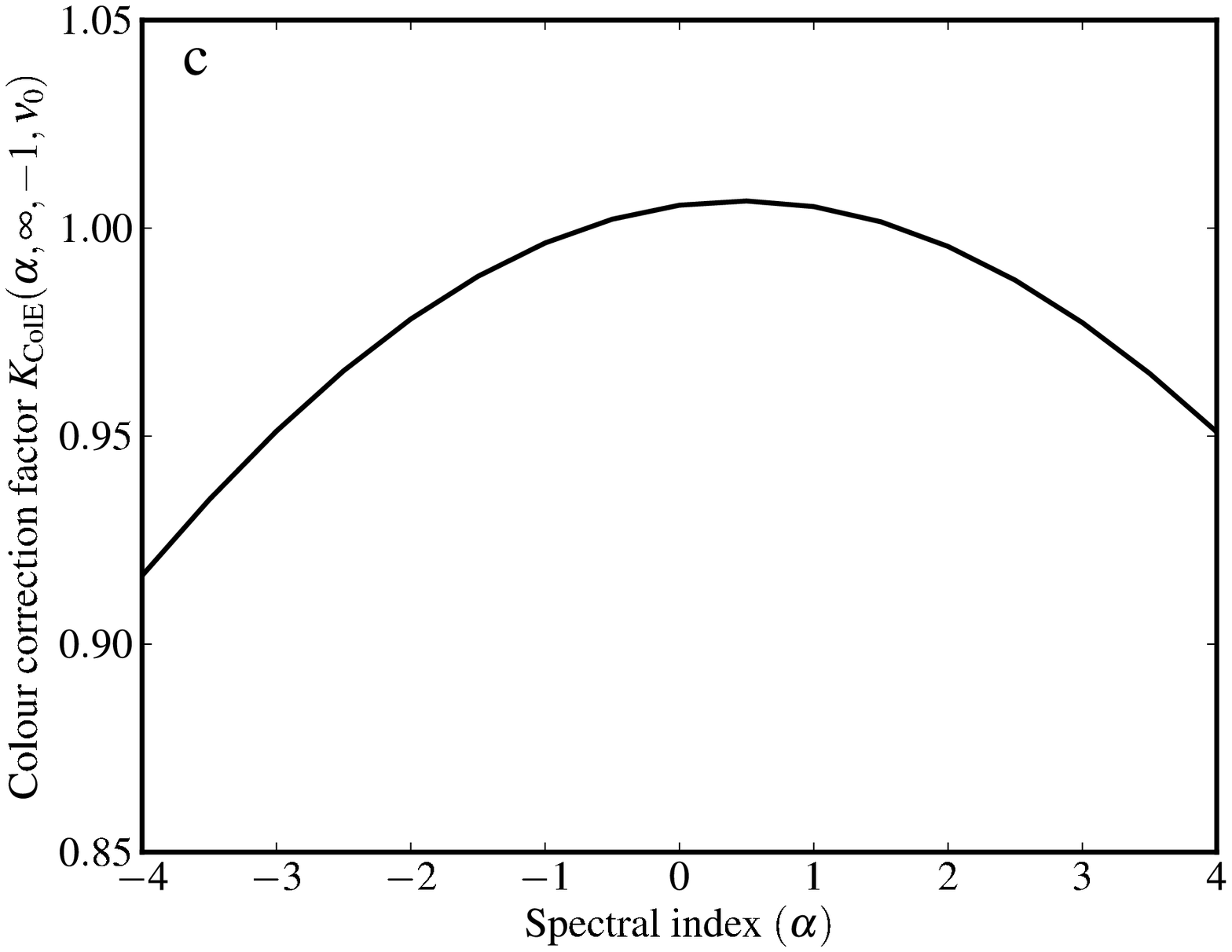}
  \caption{(a)~Point source colour correction factor vs. power-law
    source spectral index for an absorber-coupled system ($\alpha_0 =
    -1$, $\nu_0$ at band centre, pixel side = $0.5\ld$ at band centre)
    with $R=3$ (solid line), 5 (dashed line) 10 (dotted line) and 2
    (dot-dashed line). (b)~The same with $R=3$ and nominal frequency
    at the band centre (solid line), shifted by 3\% of the bandwidth
    towards the low-frequency edge (dashed line), and 3\% towards the
    high-frequency edge (dotted line). (c)~Extended source colour
    correction conversion factor vs. assumed source spectral index for
    an absorber-coupled system ($R=3$, $\alpha_0=-1$, $\nu_0$ at band
    centre, pixel side = $0.5\ld$ at band centre; fully extended
    source).}
  \label{fig:abs_uni}
\end{figure}

The results for the absorber-coupled case are shown in
\figref{fig:abs_uni}. The point source colour correction factor for an
assumed power law spectrum, $\kcolp(\alpha,-1,\nu_0)$, is shown in
\figref{fig:abs_uni}a, for a resolution of $R = 3$ and also for
narrower passbands with $R = 5$ and $R=10$, and a broader band with $R
= 2$ (in the case of an absorber coupled detector, such a broad
bandwidth is easily achievable, unlike the feedhorn case). As with the
feedhorn-coupled case (\figref{fig:ant_uni}a), depending on the source
spectral index, colour corrections on the order of 10\% can be needed
for $R\sim3$, with significantly smaller corrections for narrower
bands.  For a broader passband, the colour correction can be much
larger: up to 20\% for $R\sim2$.  Changing the nominal frequency at
which results are quoted also changes the correction significantly:
\figref{fig:abs_uni}b shows the effect of shifting the nominal
frequency in either direction by 3\% of the bandwidth. The extended
source colour correction conversion factor,
$\kcole(\alpha,\infty,-1,\nu_0)$ is plotted vs. assumed source
spectral index for the absorber-coupled case with $R = 3$ in
\figref{fig:abs_uni}c.

For either of the two architectures, the exact shapes and positions of
the correction curves will depend on the details of the passbands,
feed antenna or pixel size, and choice of nominal frequency, and they
must be computed explicitly for a given instrument.

%
%

\section{Application to the \spire\ photometer}
\label{sec:spire}

\begin{figure}
  \centering
  \includegraphics[width=0.45\textwidth]{{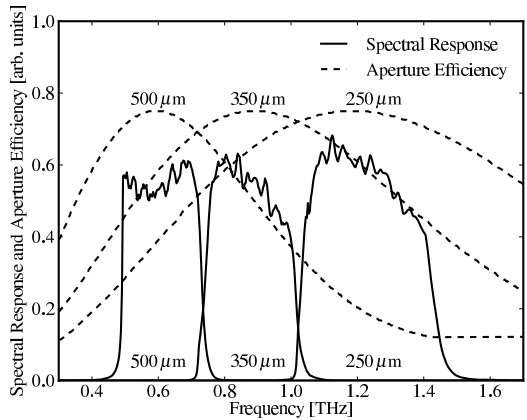}}
  \caption{Spectral response functions and aperture efficiencies as a
    function of frequency for the three SPIRE photometer bands.  Note
    that the vertical scale is irrelevant to the computations in this
    paper as all relevant parameters involve ratios.}
  \label{fig:rsrf}
\end{figure}

The calibration scheme described above has been applied to the
\spire\ camera \citep{Griffin2010_SPIRE}, and full details are given
in the SPIRE Observers' Manual\footnote{The SPIRE Observer's Manual is
  available at\\http://herschel.esac.esa.int/Docs/SPIRE/pdf/spire
  om.pdf}. Here we summarise the relevant instrument properties and
the main results. The SPIRE photometer SRFs are shown in
\figref{fig:rsrf} along with the aperture efficiency functions.  These
profiles represent the instrument transmission as a function of
frequency. The diameters of the feedhorns are sized to correspond to
$2\ld$ at wavelengths of 250, 333, and 500\,\mic, chosen to provide
high on-axis aperture efficiency for all three bands (and to ensure
that a significant subset of the detectors in the three arrays overlap
on the sky). The SPIRE bands have resolutions of $R=2.5-3.2$,
  and so are similar to the $R=3$ example discussed in
  Section~\ref{sec:resuniform}.

\subsection{Point source calibration}
\label{sec:spire_ptsrc}

The SPIRE photometer flux calibration scheme is based on the use of
Neptune as the primary calibration source, using the ESA-4
model spectrum of \citet{Moreno2012}, and the pipeline is based on the
assumption of observations of a point source.  The pipeline conversion
from detector signal to flux density essentially involves taking the
ratio of the source signal to the Neptune signal and multiplying by
the Neptune calibration flux density calculated (for the particular
time of its observation) using \eqref{eq:scal}.  Standard wavelengths
of 250, 350 and 500\,\mic\ are adopted at which the pipeline-produced
monochromatic flux densities are derived, using the nominal spectral
index $\alpha_0 = -1$, and the conversion factor
$\kmonp(-1,\nu_0)$ is applied in the pipeline.  Based on the SPIRE
SRFs and aperture efficiencies, the values of $\kmonp(-1,\nu_0)$,
given by \eqref{eq:spip}, are (1.0102, 1.0095, 1.0056) for the (250,
350, 500)\,\mic\ bands.  These numbers are not particularly sensitive
to the exact shape of the SRF or the aperture efficiency function -
for instance, assuming a flat SRF and constant aperture efficiency
across the band would give values of (1.0121, 1.0132, 1.0065).

\begin{figure}
  \centering
  \includegraphics[width=0.45\textwidth]{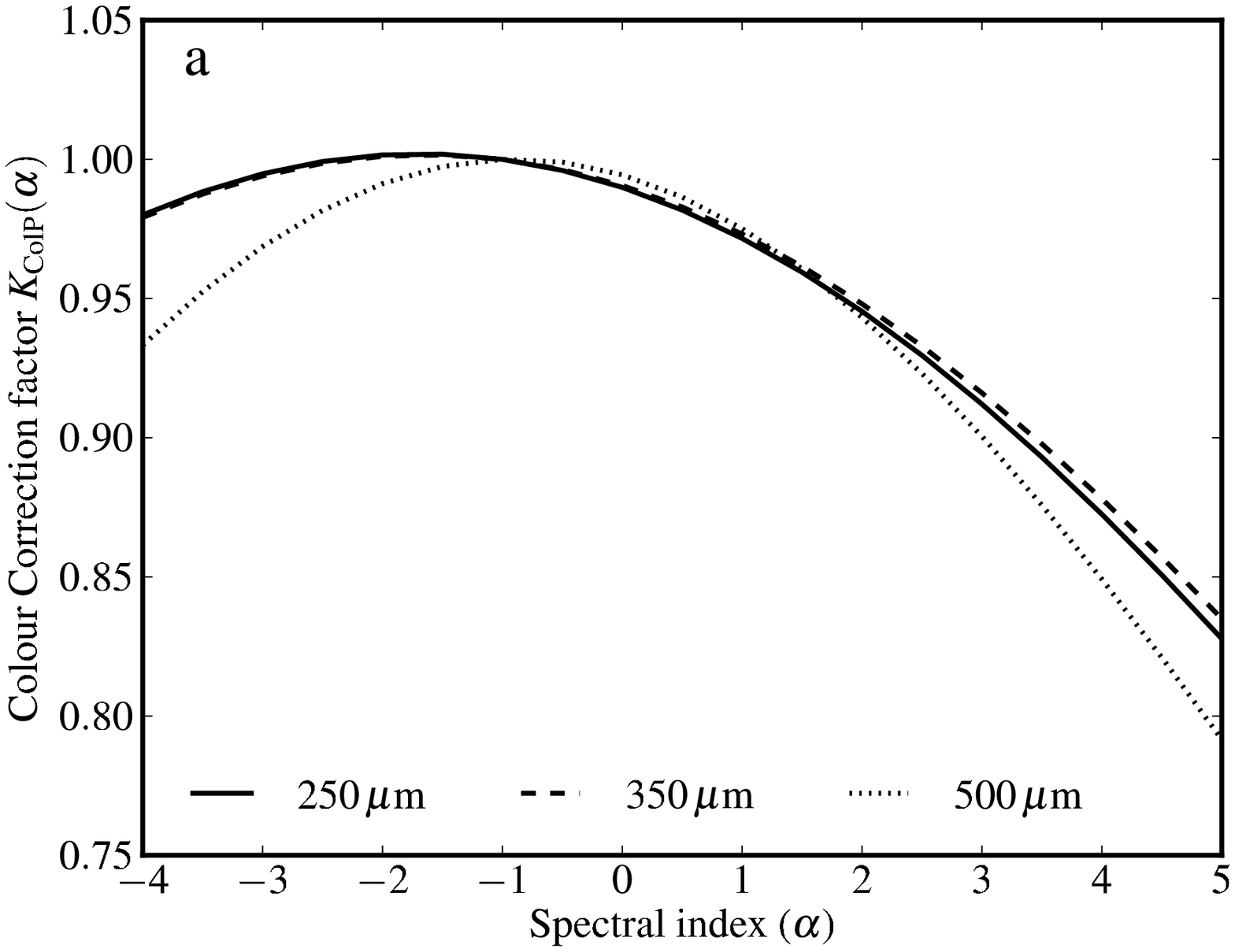}
  \includegraphics[width=0.45\textwidth]{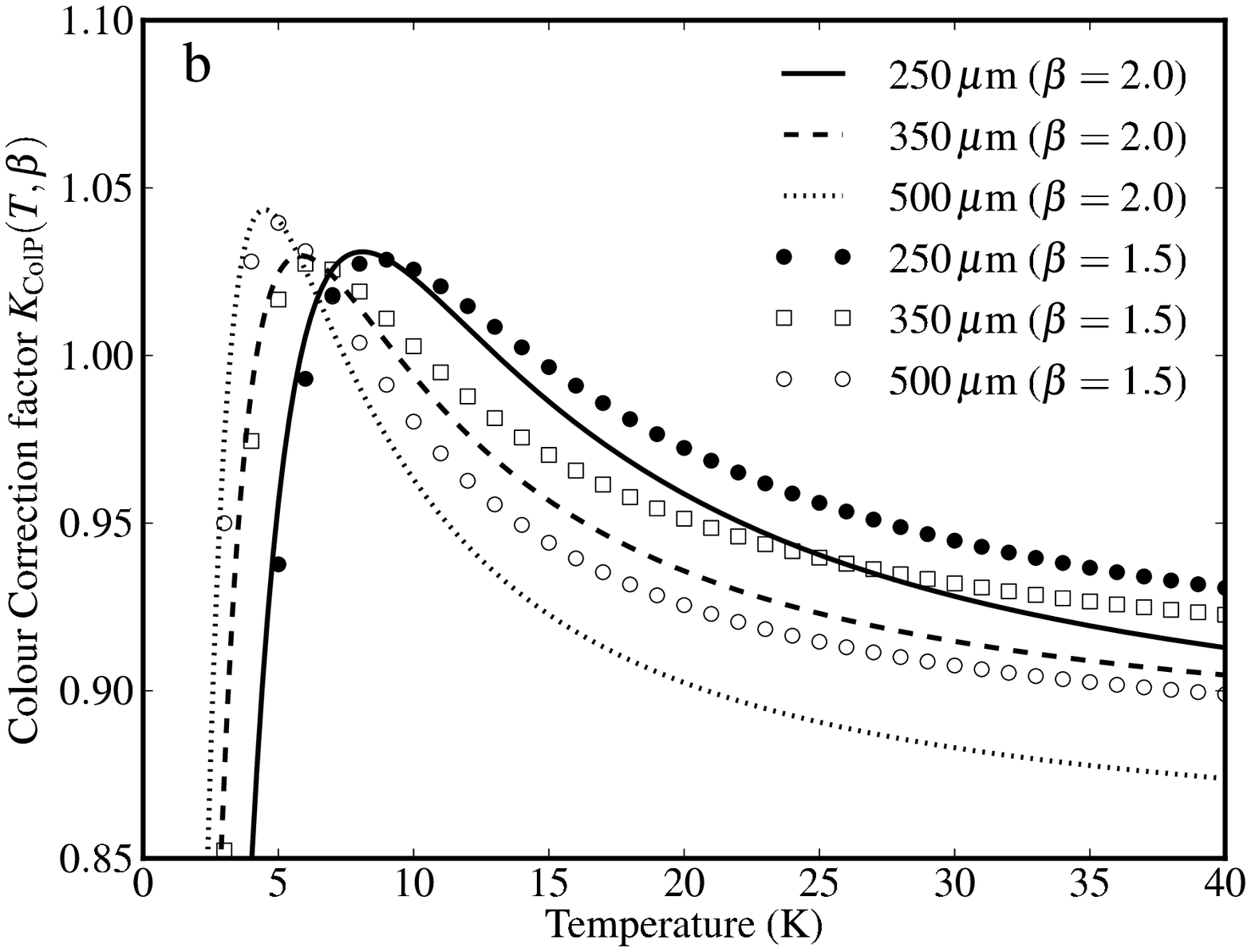}
  \caption{SPIRE point source colour correction parameter assuming (a)
    a power law source spectrum, showing $\kcolp(\alpha,-1,\nu_0)$
    vs.~assumed source spectral index, and (b)~a modified black body spectrum,
    showing $\kcolp(T,\beta,-1,\nu_0)$, vs.~source temperature for
    $\beta=2$ (lines) and 1.5 (symbols).}
  \label{fig:spire_kcolp}
\end{figure}

The SPIRE point source colour correction factors for an assumed
power law spectrum, $\kcolp(\alpha,-1,\nu_0)$, as computed using
\eqref{eq:kcolp_a}, is shown in \figref{fig:spire_kcolp}a.  Typical
dust sources have $\alpha$ in the range 1--3 in the SPIRE range,
requiring colour correction factors of a few~\%--10\%. For a $\nu^3$
source, the colour correction factors are (0.9213, 0.9248, 0.9055) for
the (250, 350, 500)\,\mic\ bands. The approximation to a power law is
reasonably accurate for a modified black body across a single SPIRE passband for
all but the lowest dust temperatures.


\figref{fig:spire_kcolp}b shows the modified black body colour correction
factors as a function of temperature up to 40\,K for two commonly
adopted values of modified black body emissivity index: $\beta=1.5$ and 2.  The
necessary correction varies significantly with source temperature.
For temperatures in the higher part of the range and $\beta=2$, the
correction factors are comparable to the power law case with
$\alpha=4$.  The lower value of $\beta$ results in smaller colour
corrections, as expected for a less steep spectrum.

As shown in Section~\ref{sec:ptsrc}, it is the value of
$\kmonp(\alpha,\nu_0)$ which converts the measured SRF-weighted flux
density to the final estimate of the source monochromatic flux
density.  It is not particularly sensitive to the assumed shape of the
SRF or aperture efficiency functions.  For instance, for a $\nu^3$
source, using $\srf$ and $\eta(\nu)$ as above gives overall factors of
$\kmonp(3,\nu_0) = (0.921, 0.925, 0.905)$ at (250, 350, 500)\,$\umu$m.
Assuming instead a top-hat SRF and constant aperture efficiency across
the band would give values of (0.917, 0.918, 0.907).  The exact shapes
of the SRFs for individual pixels vary across the detector arrays,
with the dominant effect being the movement of the band edges by
$\sim$1\% rms. Using the measured variations from the pre-launch SRF
characterisation measurements, the spread in colour correction
parameters is 1--2\%. This is in agreement for the general case
discussed in Section~\ref{sec:resuniform}, and well within the SPIRE
error budget. The effect is larger for sources with extreme spectral
indices, and is also larger for the broader 500\,\mic\ band.


\subsection{Extended source calibration}
\label{sec:spire_extsrc}

\begin{figure}
  \centering
  \includegraphics[width=0.3\textwidth]{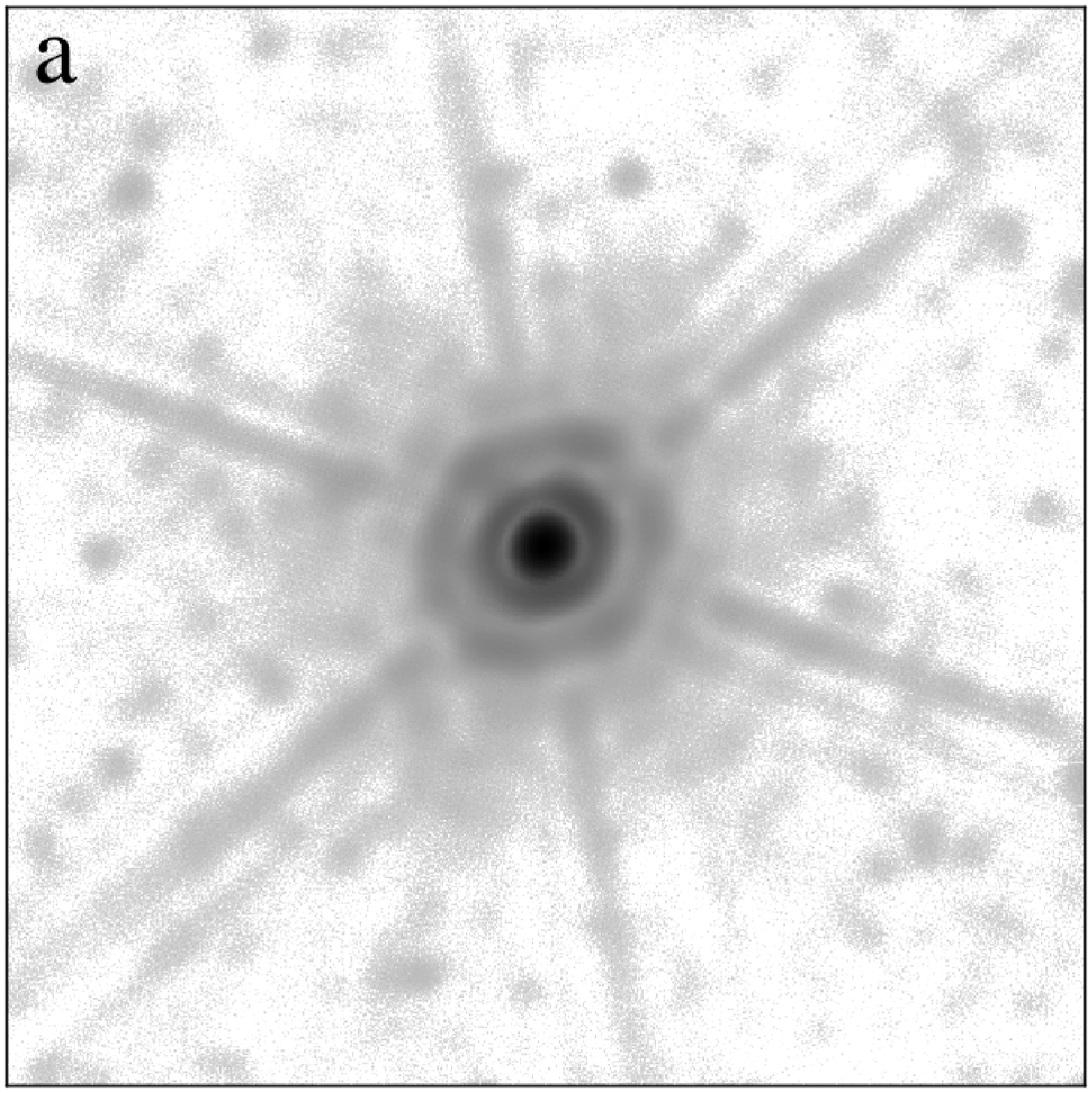}
  \includegraphics[width=0.3\textwidth]{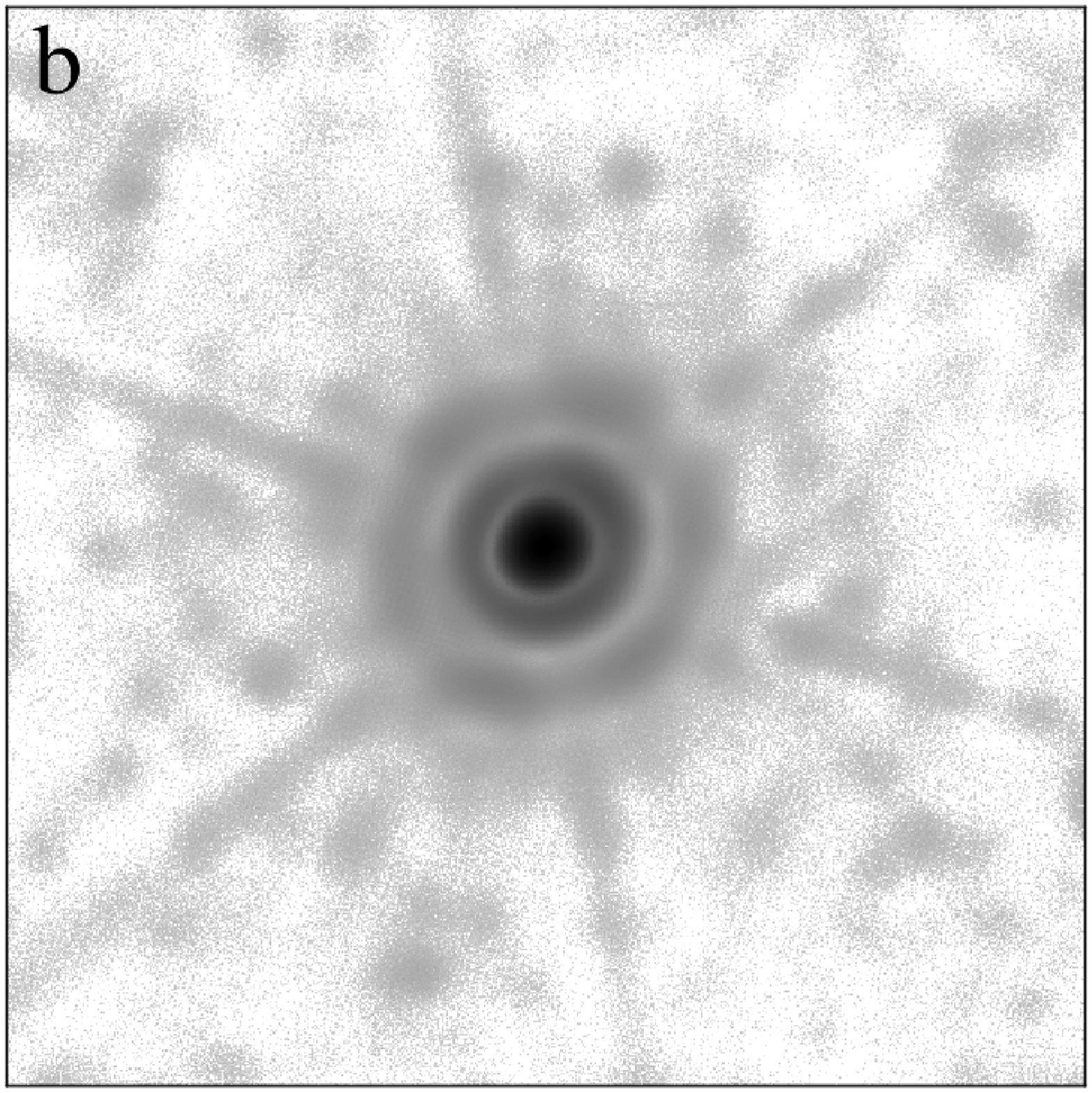}
  \includegraphics[width=0.3\textwidth]{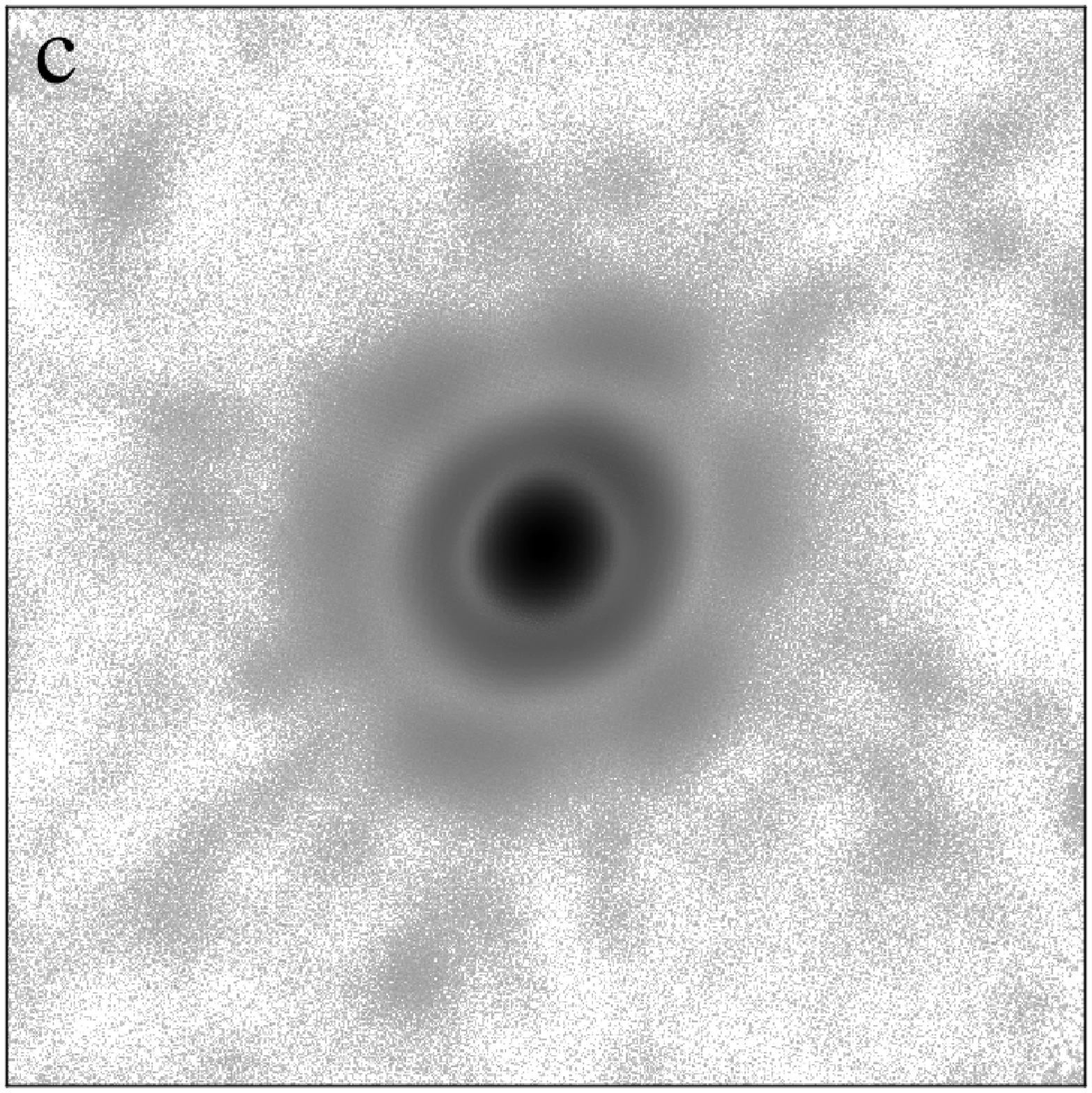}
  \caption{Broadband SPIRE beam maps at (a)~250\,\mic,
    (b)~350\,\mic, and (c)~500\,\mic.  All plots have the same linear
    spatial scale (a width of 10\,arcmin) and logarithmic colour scale
    (covering a range of [$10^{-5}$:1] relative to the peak
    value). The beam maps are of Neptune, which has a spectral index
    of $\anep = (1.29, 1.42, 1.47)$ in the SPIRE (250, 350,
    500)\,\mic\ bands. The maps show the main beam, which is broader
    at longer wavelengths, the six symmetric diffraction spikes
    due to the secondary mirror support structure. The signal-to-noise
    ratio is higher at shorter wavelengths.}
  \label{fig:beammaps}
\end{figure}

The SPIRE beams used for extended source calibration are derived from
the fine-grid scan maps of Neptune \citep{Bendo2013} shown at
1\,arcsec pixel resolution in \figref{fig:beammaps}.  To produce these
maps, the telescope was scanned across Neptune such that all
bolometers contribute equally to the map - resulting in an average
over all the detectors in a given array. Diffuse background emission
and point sources have been removed from these maps.

\subsubsection{Beam Profiles}
\label{sec:spire_beam}

\begin{figure}
  \centering
  \includegraphics[width=0.45\textwidth]{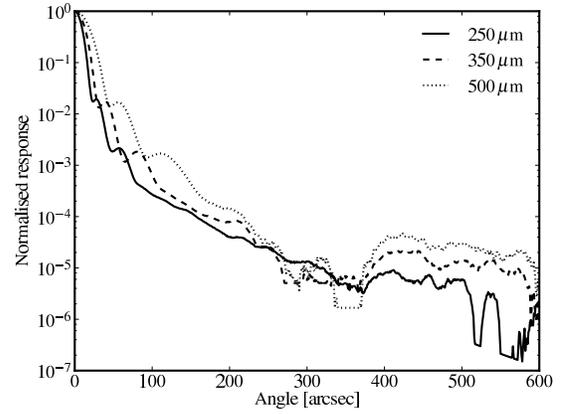}
  \caption{Azimuthally averaged measured broadband beam profiles for
    the three SPIRE bands.}
  \label{fig:beamprofs}
\end{figure}

Azimuthally averaged profiles from the measured maps are shown in
\figref{fig:beamprofs} and exhibit high signal-to-noise ratio down to
response levels $\sim10^{-5}$ and out to a radius of
$\sim250$\,arcsec.  Lower level structure, between $10^{-5}$ and
$10^{-6}$, due to the diffraction spikes produced by the secondary
support structure, is evident in the 300--500\,arcsec range with the
same basic structure for all three bands. The solid angles of these
measured band-averaged beams (computed out to a radius of 700\,arcsec)
are $\omeas = (450, 795, 1665)$\,sq.\,arcsec for (250, 350,
500)\,\mic, with uncertainties of 4\% arising from uncertainties in
the baseline level. This uncertainty is expected to be reduced in the
future by subtracting a shadow map of the same field without Neptune.

In order to create a monochromatic model for the SPIRE beams, each
azimuthally averaged beam has been split into two sections: an inner
section representing diffraction from the primary aperture,
$P_\mathrm{inner}(\theta)$, which will be scaled radially with
frequency, and an outer section which does not scale with frequency
($P_\mathrm{outer}(\theta)$). The beam profile features in this outer
section do not change significantly in angular position in the
measurements of the three bands (though they are generally broader at
longer wavelengths). This suggests that they are due to fixed objects
on the spacecraft, such as the secondary mirror support, and that they
do not vary with frequency within the bands. To derive a function for
the beam profile vs.~frequency across the band, we assume that the
measured broadband beam corresponds to the monochromatic beam at some
frequency, $\nueff$, which can be computed as shown below. The beam
profile at frequency $\nu$ within the band,
$\pmod(\theta,\nu,\nueff)$, is calculated by allowing the width of the
inner component, $P_\mathrm{inner}(\theta)$, to be scaled as
$(\nu/\nueff)^\gamma$ (where $\gamma=-0.85$, as shown in
Section~\ref{sec:antenna}), while that of the outer component
($P_\mathrm{outer}(\theta)$) is kept constant with frequency:

\begin{equation}
  \label{eq:pmod}
  \pmod(\theta,\nu,\nueff) = \mathrm{max} \left\{
    \begin{array}{l}
      P_\mathrm{inner}(\theta/(\nu/\nueff)^\gamma) \\
      P_\mathrm{outer}(\theta)
    \end{array} \right.\quad.
\end{equation}
      
Adopting $\pmod(\theta,\nu,\nueff)$ as the monochromatic beam, the
corresponding predicted broadband beam when observing Neptune
(spectral index $\anep$) is

\begin{equation}
  \label{eq:ppred}
  \ppred(\theta,\anep,\nueff) = 
  \frac{{\displaystyle \intnu \nu^{\anep} \srf \pmod(\theta,\nu,\nueff) \dd\nu}}
       {{\displaystyle \intnu \nu^{\anep} \srf \dd\nu}} \quad,
\end{equation}
with a corresponding predicted beam solid angle given by
\begin{equation}
  \label{eq:opred}
  \opred(\anep,\nueff) = \intbeam \ppred(\theta,\anep,\nueff) \, 2\upi\theta\dd\theta \quad.
\end{equation}

\begin{figure}
  \centering
  \includegraphics[width=0.45\textwidth]{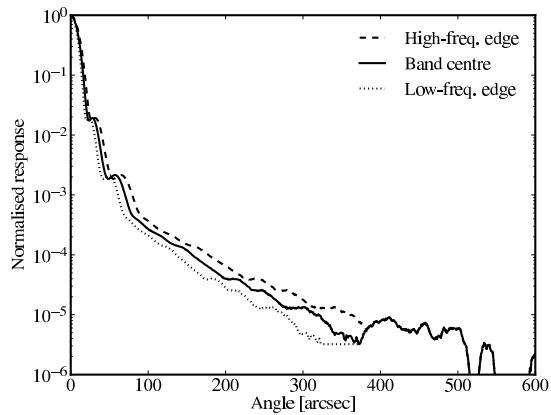}
  \caption{SPIRE monochromatic beam profile at frequency $\nueff$ for
    the 250\,\mic\ band and at the two band edges.}
  \label{fig:beamlim}
\end{figure}

\begin{figure}
  \centering
  \includegraphics[width=0.45\textwidth]{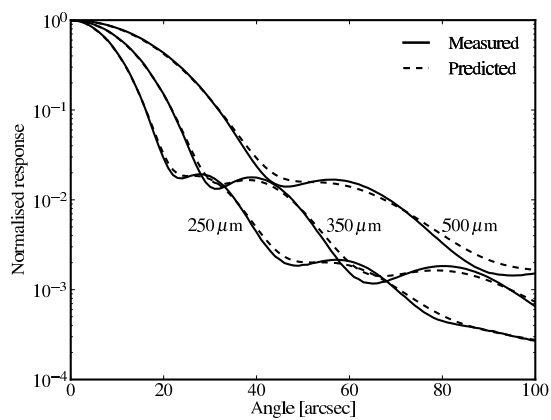}
  \caption{Measured (solid line) and predicted (dotted line) broadband
    beam profiles for the three SPIRE bands. The beam solid angles are
    equal by definition.}
  \label{fig:beamcomp}
\end{figure}

The optimum value of $\nueff$ is the value that results in this
predicted solid angle being equal to the measured value, $\omeas$.
For all three bands, frequency $\nueff$ turns out to be around 2\%
different from the nominal band frequencies, with $\nueff=(1.0133,
1.0104, 1.0177)\nu_0$ for the (250, 350, 500)\,\mic\ bands. The
corresponding monochromatic beam profile for the 250\,\mic\ band is
shown in \figref{fig:beamlim} for frequency $\nueff$, and for the two
edges of the band.  The predicted and measured broadband beam profiles
are in good agreement, as illustrated for the 250\,\mic\ band in
\figref{fig:beamcomp}.

\begin{figure}
  \centering
  \includegraphics[width=0.45\textwidth]{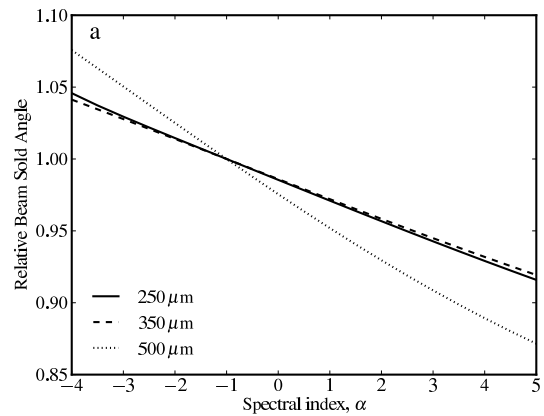}
  \includegraphics[width=0.45\textwidth]{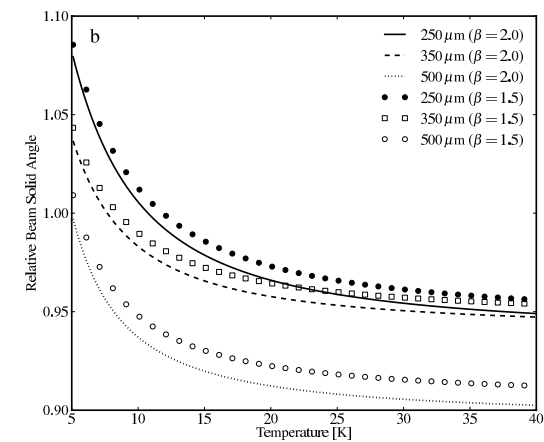}
  \caption{Effective beam solid angle, based on the modelled
      beam profiles for the three SPIRE bands, for sources with (a)~a
      range of spectral indices, and (b)~a range of temperatures for
      two values of $\beta$.}
  \label{fig:oeff}
\end{figure}

Once the value of $\nueff$ is fixed, the modelled beam profile can be
used with \eqref{eq:oeff} to produce an effective beam solid angle for
a given source spectrum, as plotted in \figref{fig:oeff} for power law
and modified black body spectra.

\subsubsection{Comparison with theoretical beam model}
\label{sec:spire_model}
An optical model of the telescope and instrument
\citep{Sibthorpe2011_BeamRelease}, covering a radius out to
300\,arcsec, produces FWHM beam widths at $\nueff$ of (17.4, 24.5,
35.0)\,arcsec, in good agreement with the measured values of (17.6,
23.9, 35.2)\,arcsec at (250, 350, 500)\,\mic\ as given in the SPIRE
Observers’ Manual.  The detailed sidelobe structure and diffraction
spikes are well reproduced by the model broadband beams, but the
modelled solid angles are 7-9\% smaller than the measured solid
angles. This may reflect contributions by additional stray light and
diffractive effects not explicitly accounted for in the model.

\subsubsection{Results for a fully extended source}
\label{sec:spire_res}

\begin{figure}
  \centering
  \includegraphics[width=0.45\textwidth]{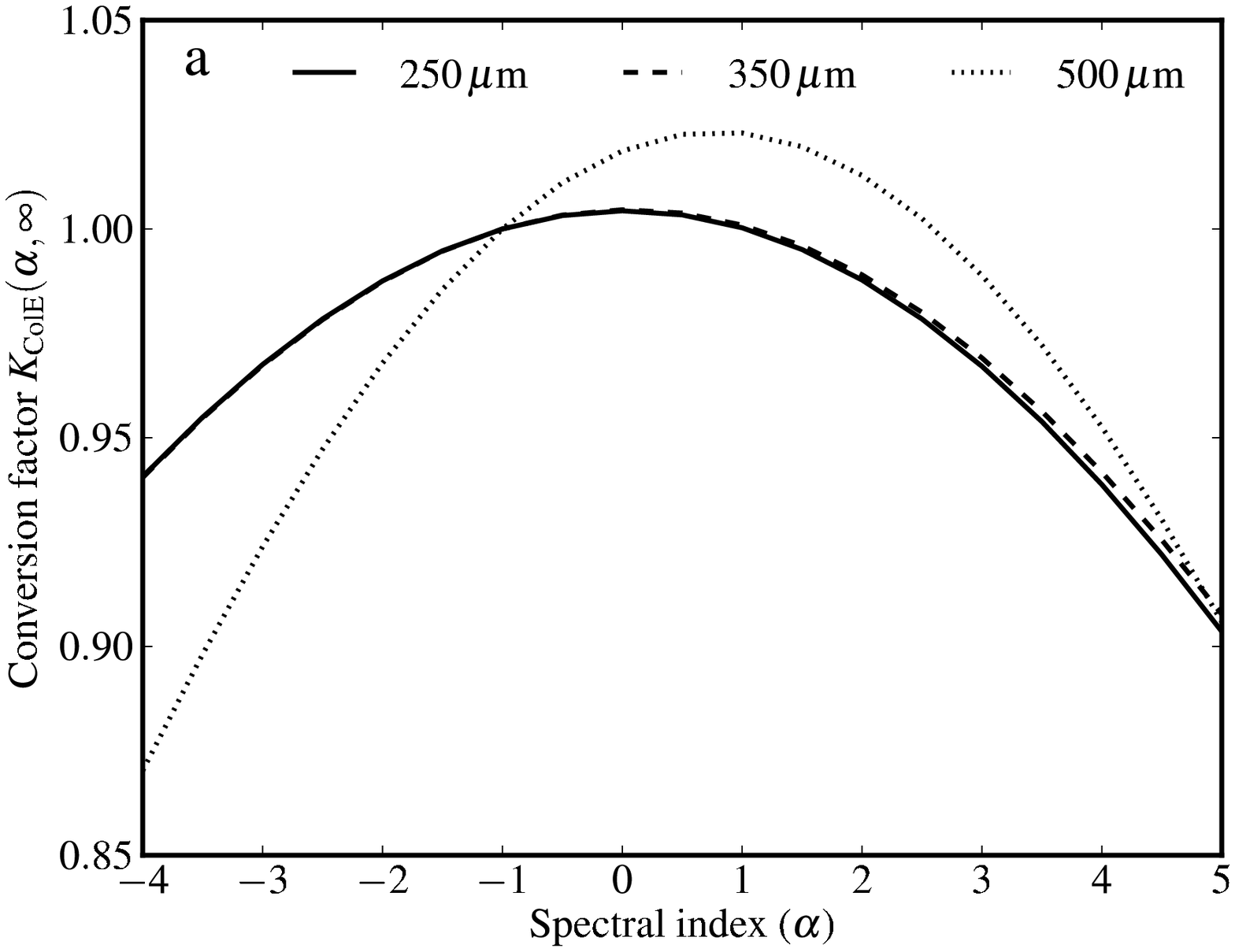}
  \includegraphics[width=0.45\textwidth]{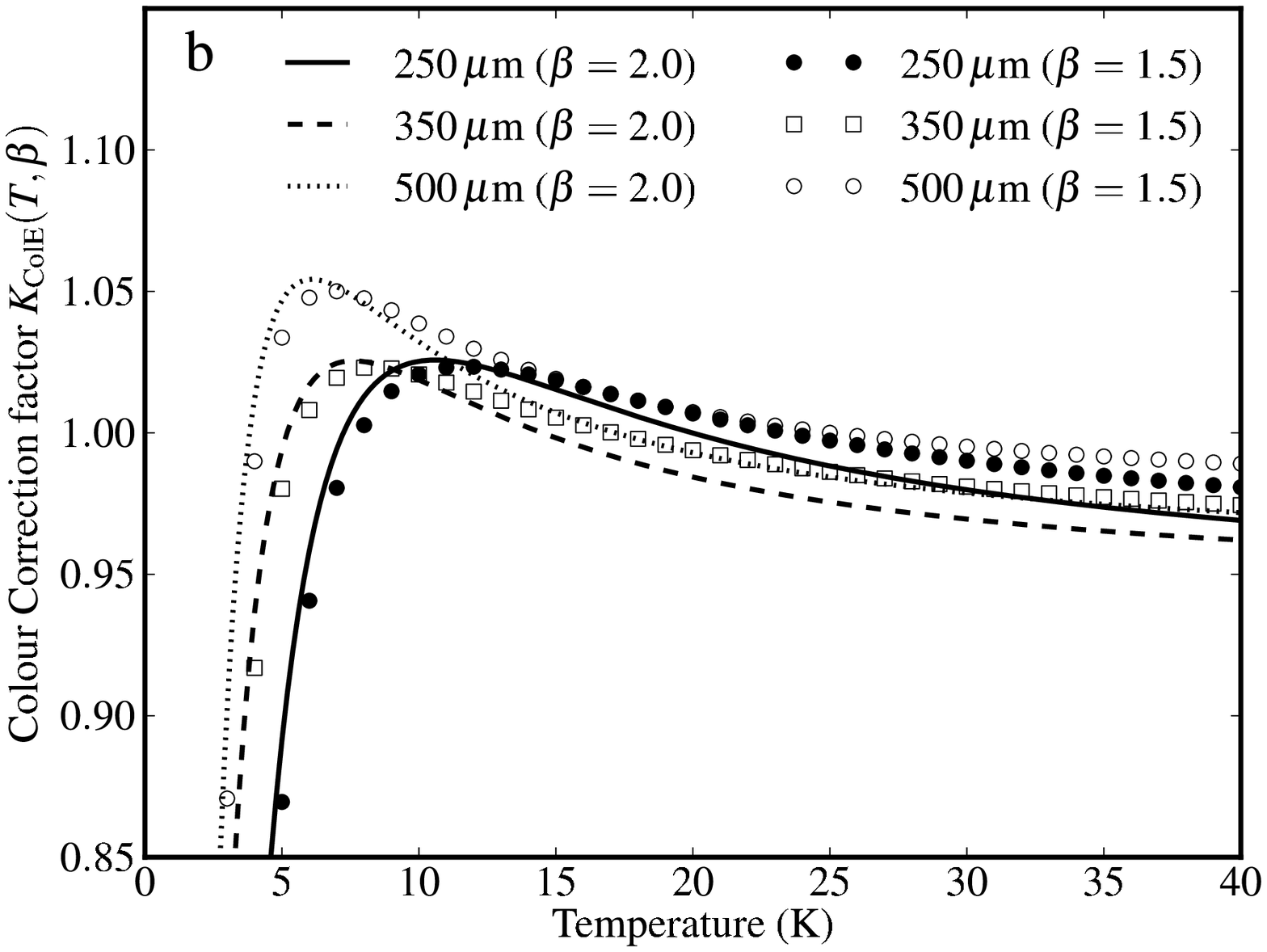}
  \caption{The colour correction factor factor for a fully-extended
    source, assuming (a)~a power law source spectrum, and (b)~a modified black
    body source spectrum with $\beta=2$ (lines) and 1.5 (symbols).}
  \label{fig:spire_kext}
\end{figure}

The factors for converting from the point source pipeline to the
extended source pipeline for SPIRE are (91.419, 51.721, 24.055)\,MJy/sr
per Jy for the (250, 350, 500)\,\mic\ bands.  The colour correction
factors for fully extended sources, $\kcole(f,\infty,-1,\nu_0)$, for
converting the standard SPIRE extended pipeline surface brightness to
the value for a fully extended source of a given spectrum, are plotted
in \figref{fig:spire_kext}a (tabulations of this and the other results
shown below for SPIRE are given in the SPIRE Observer's Manual). These
results are not sensitive to a small change in the adopted value of
$\gamma$, the power-law index of the frequency scaling of the beam
profile.  For a $\nu^3$ source, using $\gamma=0.75$ or 0.95 instead of
0.85 results in $\kcole(3,\infty,-1,\nu_0)$ changing by less than
0.1\% for all three bands.  However, assuming that the beam profile is
independent of frequency ($\gamma$ = 0) results in values which are up
to 10\% different for the three bands.

$\kcole(T,\beta,\infty,-1,\nu_0)$, the correction factor for an assumed modified black
body spectrum, is shown as a function of modified black body temperature in
\figref{fig:spire_kext}b for $\beta=1.5$ and 2.  The dependence on
$\beta$ is small, and the dependence on $T$ is also small for
temperatures above $\sim5$\,K for the 500\,\mic\ band and $\sim10$\,K
for the 250\,\mic\ band.

An earlier version of the SPIRE calibration scheme
(\citet{Swinyard2010_SPIRECal}, and SPIRE Observer's Manual v2.4 and
earlier) accounted for the variation of beam size across the band by
weighting the SRF by the square of the wavelength (equivalent to
$\gamma = 1$ and $\delta=2$), producing a larger throughput at the
longer-wavelength end of the band.  Different colour correction
factors were derived and quoted for the case of point and extended
sources, and the conversion from flux density to surface brightness
was carried out by dividing by the broadband beam area as measured on
Neptune. Compared to the new method presented here, this earlier
method produced surface brightness values which are higher by
approximately (7, 7, 12)\% at (250, 350, 500\,\mic) for a $\nu^3$
source. We note that these systematic errors were within the $\pm15$\%
quoted uncertainties.



It is also interesting to compare the results obtained for a fully
extended source using the new method and by simply dividing the
pipeline output by the measured broadband beam area.  For a source of
spectral index $\alpha$, this produces an estimate of the sky surface
brightness that departs from the true value by a factor of $G(\alpha)$ given by
\begin{equation}
  \label{eq:galpha}
G(\alpha) = \frac{\displaystyle \kmonp(\alpha,\nu_0)} {\displaystyle
  \kuni(\alpha,\nu_0) \omeas} \quad.
\end{equation}

The effect is greater for sources with spectra that are more different
from that of Neptune. For a $\nu^2$ source, $G(2) = (0.991, 0.992,
0.989)$ and for $\nu^3$, $G(3) = (0.976, 0.979, 0.968)$.  The sky
intensity, which depends on the source spectral index, is thus
underestimated by a small extent, and can be up to $\sim~3$\% at
500\,\mic\ for a $\nu^3$ source.

\subsubsection{Results for a partially extended Gaussian source}
\label{sec:spire_partial}

\begin{figure}
  \centering
  \includegraphics[width=0.45\textwidth]{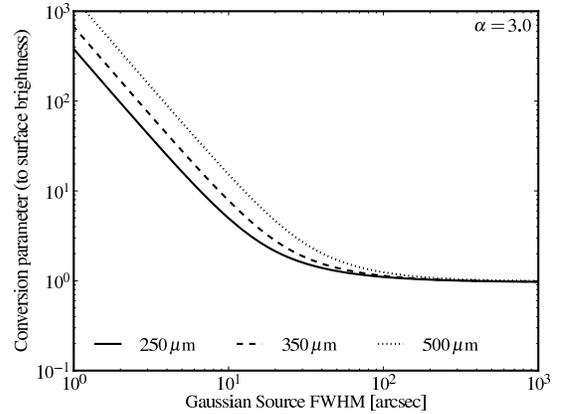}
  \caption{Conversion factor from the surface brightness produced by
    the extended source pipeline to the peak surface brightness of a
    partially-exended source, plotted against the source FWHM, for a
    source with $\alpha=3$.}
  \label{fig:spire_kextg}
\end{figure}

A partially extended source lies between the point-like and
fully-extended sources. The conversion factor from the extended
pipeline peak surface brightness (which applies to a fully extended
source, as adopted for the SPIRE extended source pipeline) to the peak
surface brightness for a partially extended source,
$\kcole(\alpha,\theta_0,-1,\nu_0)$ is plotted vs.~the source FWHM in
\figref{fig:spire_kextg} for the case of $\alpha=3$ (a typical value
for a cold dust source observed by SPIRE).  For large source widths
(i.e.~ $\theta_0 \to \infty$) it converges on the colour correction
parameters plotted in \figref{fig:spire_kext}a, which are close to
unity, since the peak value tends towards the surface brightness of of
a fully-extended source. For small sources this conversion also
includes the compensation for the fact that the flux is from a smaller
solid angle, defined by the combination of the source size and the
effective beam, and so the peak value of the true source surface
brightness must increase.

\begin{figure}
  \centering
  \includegraphics[width=0.45\textwidth]{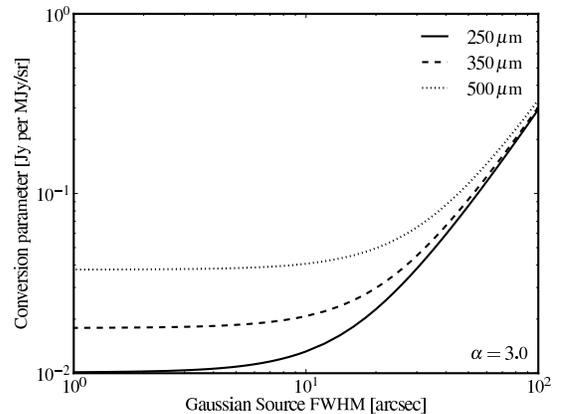}
  \caption{Conversion factor from the surface brightness produced by
    the extended source pipeline to the total flux density of a
    partially extended Gaussian source, plotted against the source
    FWHM, for the case of $\alpha=3$.}
  \label{fig:spire_kextg_tot}
\end{figure}

The total flux density of a partly extended source can be calculated
by multiplying the peak surface brightness by the effective source
area. The corresponding conversion factor between the peak pipeline
surface brightness and the total integrated flux density of the source
is plotted as a function of source FWHM in
\figref{fig:spire_kextg_tot}. For very small sources (i.e.~$\theta_0
\to 0$), this converges on a value that returns the colour-corrected
point source flux. For large sources the total flux density increases
with the source area.

\section{Conclusions}
\label{sec:conc}
We have developed a methodology for flux calibration of
FIR-submillimetre observations made with broadband instruments, using
either antenna-coupled or absorber-coupled detectors. It takes into
account the variation of beam width and aperture efficiency across the
photometric band.  It can accommodate arbitrary source SEDs and, in
the case of extended emission, arbitrary source surface brightness
profiles.  Accurate knowledge of the instrument properties and of the
beam profile is needed to ensure that extended emission can be
calibrated with respect to a point source standard.

Application of this scheme to the case of the \spire\ photometer
produces results which are a few percent higher than those obtained by
ignoring the fact that the beam profile varies across the
passband. Although not large, these systematic effects are comparable
to the absolute uncertainties of the primary calibrator, and so need
to be understood and eliminated from the overall error budget.

Similar considerations will apply to other broadband photometric
instruments (space-borne in the submillimetre region or ground-based
at longer wavelengths).  Additional practical aspects of SPIRE flux
calibration are covered in a companion paper Bendo et al. (in
preparation) and in the SPIRE Observers' Manual.

\section{Acknowledgements}
\label{sec:acknowledgement}
{\em Herschel} is an ESA space observatory with science instruments
provided by European-led Principal Investigator consortia and with important
participation from NASA.

SPIRE has been developed by a consortium of institutes led by Cardiff
Univ. (UK) and including: Univ. Lethbridge (Canada); NAOC (China);
CEA, LAM (France); IFSI, Univ. Padua (Italy); IAC (Spain); Stockholm
Observatory (Sweden); Imperial College London, RAL, UCL-MSSL, UKATC,
Univ. Sussex (UK); and Caltech, JPL, NHSC, Univ. Colorado (USA). This
development has been supported by national funding agencies: CSA
(Canada); NAOC (China); CEA, CNES, CNRS (France); ASI (Italy); MCINN
(Spain); SNSB (Sweden); STFC, UKSA (UK); and NASA (USA).

This research made use of \textsc{APLpy}, an open-source plotting package for
\textsc{Python} hosted at \texttt{http://aplpy.github.com}

\bibliographystyle{mn2e}
\bibliography{abbrev_cen,fluxcal}


\section*{Appendix}
\label{sec:appendix}
A list of symbols used in the paper is given in \tabref{tab:symbols}.
\begin{table*}
    \caption{List of symbols}
    \label{tab:symbols}
    \begin{tabular}{r p{0.8\textwidth}}
      {\bf Symbol} & {\bf Definition} \\

      $\mathcal{B}(\nu,T)$ & Planck function for temperature, $T$, and frequency, $\nu$ \\
      $B(\nu,\theta,\phi)$ & The beam response as a function of position $(\theta,\phi)$ and frequency, $\nu$ \\
      $\srf$ & Spectral response function (SRF) of a photometric band in terms of frequency, $\nu$\\
      $f(\nu,\nu_0)$ & Source spectrum normalized to the flux density at frequency $\nu_0$\\
      $g(\theta,\theta_0)$ & Radial intensity profile, as a function of radial offset angle $\theta$, of a source with a width characterized by parameter $\theta_0$ \\
      $G(\alpha)$ & Factor by which a na\"\i ve approach to extended source calibration results is an incorrect estimate of the sky surface brightness \\
      $\ipip(\alpha_0,\nu_0)$ & Monochromatic peak surface brightness of a fully extended source produced by the extended source pipeline, assuming a spectral index $\alpha_0$.\\ 
      $I(\nu,\theta)$ & Radial source surface brightness profile as a function of  frequency $\nu$ and radial offset angle $\theta$. \\
      $I(\nu,\theta,\phi)$ & Radial source surface brightness as a function of  frequency $\nu$ and angular position $(\theta,\phi)$. \\
      $K_\mathrm{Beam}(\theta_\mathrm{P},\theta_\mathrm{Beam})$ & Beam correction factor for a Gaussian main beam coupling to a uniform disk source \\
      $\kcole(f,g,\alpha_0,\nu_0)$ & Factor to convert the monochromatic pipeline extended source surface brightness, $\ipip(\nu_0)$, to the true peak surface brightness of a source with spectrum $f(\nu,\nu_0)$ and spatial variation $g(\theta,\theta_0)$. \\
      $\kcolp(f,\alpha_0,\nu_0)$ & Colour correction factor to convert the monochromatic pipeline point source flux density at nominal frequency $\nu_0$, $\spip(\nu_0)$, to that corresponding to a different assumed source spectrum $f(\nu,\nu_0)$.\\
      $\kmone(f,g,\nu_0,\theta_0)$ & Factor to convert SRF-weighted flux density, $\smeas$, to monochromatic surface brightness, $I(\nu)$, at frequency $\nu$ for an extended source with spectrum $f(\nu,\nu_0)$ and spatial variation $g(\theta,\theta_0)$. \\
      $\kmonp(f,\nu_0)$ & Factor to convert SRF-weighted flux density, $\smeas$ to monochromatic flux density, $S(\nu_0)$, at frequency $\nu_0$ for a point source with a spectrum given by $f(\nu,\nu_0)$. \\
      $\kuni(f,\nu_0)$ & The conversion parameter for a fully extended source, i.e., $\kmone(f,g,\alpha_0,\nu_0)$ with $\theta_0=\infty$ and therefore $g(\theta,\theta_0)\equiv 1$ \\
      $P(\nu,\theta)$ & Normalised, azimuthally-averaged beam profile as a function of radial offset angle $\theta$ and frequency $\nu$ \\
      $P(\nu,\theta,\phi)$ & Normalised beam response as a function of orthogonal offset angles $\theta$ and $\phi$, and frequency $\nu$ \\
      $P_\mathrm{inner}(\theta)$, $P_\mathrm{outer}(\theta)$ & Inner and outer portions of the monochromatic beam profile. The inner portion is scaled with frequency, while the outer portion is not. \\
      $\pmeas(\theta,\alpha)$ & Broadband beam profile measured on a source of spectral index $\alpha$ \\
      $P_\mathrm{mod}(\theta,\nu,\nueff)$ & Modelled monochromatic beam profile at frequency $\nu$, assuming the measured beam is equivalent to the monochromatic beam at frequency $\nueff$\\
      $\ppred(\theta,\alpha,\nueff)$ & Predicted broadband beam profile for a source spectral index $\alpha$ and effective frequency $\nueff$\\
      $R$ & Passband width relative to the central frequency \\
      $S(\nu)$ & Source flux density at frequency $\nu$ \\
      $\overline{S}_\mathrm{C}$ & SRF-weighted flux density for a calibration source \\
      $\smeas$ & SRF-weighted flux density for an observed source \\
      $\spip(\alpha_0,\nu_0)$ & Monochromatic flux density at frequency $\nu_0$ produced by a pipeline that assumes point source calibration and a source spectral index of $\alpha_0$ \\
      $T$ & Black body or modified black body temperature \\
      $y(\nu)$ & Monochromatic flux density at frequency $\nu$ for an extended source with surface brightness $I(\nu,\theta,\nu)$, integrated over a monochromatic beam with response $B(\nu,\theta,\phi)$ \\
      $y^\prime(\nu,\theta_0)$ & Normalised monochromatic flux density at frequency $\nu$ for an extended source with radial intensity profile $g(\theta,\theta_0)$, integrated over a monochromatic beam with radial profile $P(\nu,\theta)$ \\
      \hline
      $\alpha$ & Astronomical source power law spectral index \\
      $\alpha_0$ & Nominal source spectral index for which SPIRE flux densities are quoted \\
      $\anep$ & Spectral index adopted for Neptune when used as a SPIRE photometric calibrator \\
      $\beta$ & modified black body emissivity index such that emissivity $\propto \nu^\beta$ \\
      $\gamma$ & Power law index for adopted variation of main beam FWHM with frequency \\
      $\delta$ & Power law index for adopted variation of beam solid angle with frequency \\
      $\eta(\nu)$ & Aperture efficiency (total power coiupled to detector from an on-axis source) as a function of frequency, $\nu$\\
      $\theta$ & Radial offset angle from the centre of the beam \\
      $\theta_0$ & Width parameter of a source with a circularly-symmetric intensity profile \\
      $\theta_\mathrm{Beam}$ & Beam FWHM used for calibration \\
      $\theta_P$ & Angular radius of observed planetary disk used as a calibrator\\
      $\lambda$, $\nu$ & Radiation wavelength and frequency \\
      $\Delta\lambda$, $\Delta\nu$ & Bandwidth of photometer passband in terms of wavelength and frequency\\
      $\nu_0$ & Nominal frequency at which monochromatic source flux density is to be quoted\\
      $\nueff$ & Effective frequency at which the monochromatic beam profile is assumed to be equal to the measured beam profile, and constrained such that $\opred$ is equal to $\omeas$ \\
      $\phi$ & Azimuthal offset angle relative to the beam centre, used in non-circularly symmetric cases \\
      $\Omega(\nu)$ & Monochromatic beam solid angle at frequency $\nu$ \\
      $\oeff(f)$ & Effective beam solid angle for observations of a source with spectrum $f(\nu,\nu_0)$ \\
      $\omeas(\alpha)$ & Broadband beam solid angle as measured on a point source of spectral index $\alpha$\\
      $\onorm(\nu,\nu_0)$ & Monochromatic beam solid angle at frequency $\nu$ normalised to the value at $\nu_0$ \\
      $\opred(\alpha,\nueff)$ & Predicted beam solid angle for a source with spectral index $\alpha$, assuming an effective frequency $\nueff$\\
    \end{tabular}
\end{table*}


\label{lastpage}

\end{document}